\documentclass[10pt,superscriptaddress,amsmath,amssymb,aps,prx,twocolumn,a4paper,floatfix,longbibliography]{revtex4-1}

\usepackage[version=3]{mhchem}
\usepackage{graphicx,times,bm,amssymb,amsmath,booktabs}
\usepackage{physics,subfigure,makecell,hyperref,multirow,color}
\usepackage{times,color}
\usepackage{amsfonts}
\usepackage{mathrsfs}

\newcommand{\matheq}[2]{
  \begin{equation}  \label{#1}
        #2
   \end{equation}
  }

\hypersetup{colorlinks,allcolors=black}
\def\be{\begin{equation}}       \def\ee{\end{equation}}
\def\bea{\begin{eqnarray}}      \def\eea{\end{eqnarray}}

\usepackage{graphicx}
\usepackage{epstopdf}

\begin{document}
\title{Optimal coloring and strain-enhanced superconductivity in Li$_n$B$_{n+1}$C$_{n-1}$}

\author{Yuhao Gu}
%\email{guyuhao@iphy.ac.cn}
\affiliation{School of Mathematics and Physics, University of Science and Technology Beijing, Beijing 100083, China}

\author{Jiangping Hu}\email{jphu@iphy.ac.cn}
\affiliation{Institute of Physics, Chinese Academy of Sciences, Beijing 100190, China}
\affiliation{Kavli Institute of Theoretical Sciences, University of Chinese Academy of Sciences,
Beijing, 100190, China}
\affiliation{New Cornerstone Science Laboratory, Beijing 100190, China}

\author{Hong Jiang}
\email{jianghchem@pku.edu.cn}
\affiliation{Beijing National Laboratory for Molecular Sciences, College of Chemistry and Molecular Engineering, Peking University,Beijing 100871, China}

\author{Tao Xiang}%\email{txiang@iphy.ac.cn}
\affiliation{Institute of Physics, Chinese Academy of Sciences, Beijing 100190, China}
\affiliation{Beijing Academy of Quantum Information Sciences,
Beijing, 100193, China}
\affiliation{Kavli Institute of Theoretical Sciences, University of Chinese Academy of Sciences,
Beijing, 100190, China}

\begin{abstract}
 Boron-rich lithium borocarbides are promising candidates for phonon-mediated high-temperature superconductors due to their metallic $\sigma$-bonding electrons. Here, we use the cluster expansion method to identify energetically stable configurations (colorings) of \ce{Li2B3C} and \ce{Li3B4C2}, which are characterized by a distinctive pattern of alternating B-B and B-C zigzag chains. Surprisingly, the optimal configuration of \ce{Li2B3C} exhibits an extremely low superconducting transition temperature of $T_c < 0.03$ K, which is attributed to the suppression of deformation potentials near the Fermi level caused by the specific electron filling of B-B zigzag chains. However, the $\sigma$-bonding electrons at the Fermi level are highly sensitive to external strain or pressure. Specifically, applying a -5\% compressive uniaxial strain can significantly enhance the electron-phonon coupling and the Eliashberg spectral function, boosting up $T_c$ to 37 K. This work not only presents a novel strategy for achieving phonon-mediated high-temperature superconductivity in \ce{Li_nB_{n+1}C_{n-1}} compounds but also provides valuable insights into the complex interplay between electronic structure and superconducting interaction.

\end{abstract}

\maketitle

\section{Introduction}

 The allure of high-temperature superconductivity continues to fascinate the scientific community worldwide. The metallization of strong chemical bonding electrons is considered a pivotal factor in the formation of high-$T_c$ superconducting pairings in copper oxides \cite{bednorz1986possible}, \ce{MgB2} \cite{nagamatsu2001superconductivity} and hydrogen-rich materials under ultrahigh pressure \cite{drozdov2015conventional}.

 The search for new superconductors with high transition temperature ($T_c$) has attracted tremendous interest since the discovery of superconductivity. Inspired by the discovery of high-$T_c$ superconductivity in cuprates and \ce{MgB2}, the metallization of $\sigma$-bonding electrons has been proposed as a universal guiding principle \cite{gao2015chinese} for discovering high-temperature superconductors. This concept has been widely applied to numerous phonon-mediated superconductors \cite{mgb2_pickett, gao2015prediction, deng2003mgb2flat,deng2005flat,lee2004superconductivity, drozdov2015conventional, peng2017hydrogen,gu2021bacus2,geng2023B3C3,geng2024superconductivity}.

 A $\sigma$-bond is the strongest covalent bond formed by the direct overlap of two atomic orbitals, each housing a single electron. In contrast, a $\sigma$-antibonding state typically corresponds to a repulsive interaction in a diatomic molecule. However, in solid materials,  $\sigma$-antibonding states can also play a crucial role by contributing to the attractive interaction that binds electrons together, as highlighted in Ref. \cite{ZhangJianFeng2023}. Electronic bands originating from hybridized $\sigma$-bonding or $\sigma$-antibonding orbitals are often situated far below or above the Fermi level.

 As discussed in Refs. \cite{gao2015chinese, ZhangJianFeng2023}, there are a number of ways to metalize $\sigma$-hybrized electrons. One way of inducing metallization involves the application of external pressure, as demonstrated in the hydrogen-rich high-$T_c$ superconductor, SH$_3$, under ultra-high pressure \cite{drozdov2015conventional}. This external pressure, combined with the chemical forces exerted on hydrogen atoms by neighbouring atoms, can stabilize crystal structures. It can also alter the electron filling factor, raising the Fermi level crossing the bands formed by the antibonding-$\sigma$ electrons.  An alternative approach is through the exploitation of the crystal field effect, as exemplified by magnesium diboride \ce{MgB2} \cite{mgb2_mazin, mgb2_pickett, mgb2_xxx, gao2015chinese}. In \ce{MgB2}, the crystal structure consists of alternating layers of magnesium atoms arranged in a triangular lattice and boron atoms in a honeycomb lattice. The crystal field effect induces a strong attraction between the \ce{Mg^{2+}} ions and the $\pi$ electrons in the boron layers, lowering the energy of the bands associated with the hybridized $\pi$ orbitals. This interaction elevates the energy of the $\sigma$-hybridized boron orbitals, bringing them into intersection with the Fermi level. A more general and straightforward strategy is to manipulate the Fermi level by doping. Copper-oxide high-$T_c$ superconductors are typical examples of this type.

 The superconductivity in magnesium diboride \ce{MgB2} has spurred interest in  exploring other superconductors that exhibit metallic $\sigma$-bonding electrons \cite{rosner2002prediction, dewhurst2003first, miao2013LiB1.1C0.9, bazhirov2014Li4B5C3,gao2015prediction, li2018LiB1+xC1-x,quan_Pickett_2020li2Xbc3,liu2024three,yu2024nontrivial,pickett2006design,pickett2017revolution}. Lithium borocarbide LiBC, isoelectronic with \ce{MgB2}, has garnered significant attention in this context \cite{rosner2002prediction, dewhurst2003first}. While pure LiBC is a band insulator \cite{fogg2003libc, karimov2004libc}, theoretical studies suggest that introducing lithium deficiencies to create hole doping in \ce{Li_{1-x}BC} could lead to superconductivity at relatively high critical temperatures \cite{rosner2002prediction, dewhurst2003first}. However, experimental attempts to synthesize \ce{Li_{1-x}BC} with finite lithium deficiencies have not been successful in stabilizing the expected crystal structure \cite{bharathi2002synthesis,fogg2003libc, fogg2003synthesis, souptel2003synthesis, zhao2003synthesis, nakamori2004synthesis, fogg2006chemical}, prohibiting the expected metallization of the $\sigma$-bonding electrons \cite{fogg2006chemical}.

 To mitigate the lattice instability caused by Li deficiencies, Gao et al. suggested partially substituting C atoms with B atoms in LiBC  \cite{miao2013LiB1.1C0.9, gao2015chinese}. Through first-principles calculations, they identified two potential superconducting compounds within the \ce{Li_nB_{n+1}C_{n-1}} series, \ce{Li2B3C} ($n=2$) and \ce{Li3B4C2} ($n=3$), which were predicted to exhibit critical transition temperatures exceeding 50 K \cite{gao2015prediction}. Later on, similar proposals were made by several other groups \cite{miao2013LiB1.1C0.9, bazhirov2014Li4B5C3,  li2018LiB1+xC1-x}.

 However, determining the optimal configuration of boron and carbon atoms in \ce{Li_nB_{n+1}C_{n-1}}, known as the "coloring problem" \cite{burdett1985coloring, miller1998coloring}, remains a challenge. Gao et al. \cite{gao2015prediction} and Bazhirov et al. \cite{bazhirov2014Li4B5C3} investigated the electronic properties of these compounds using hypothetical crystal structures. Other studies \cite{miao2013LiB1.1C0.9, li2018LiB1+xC1-x} model the substitution effect by employing the virtual crystal approximation.

 In this work, we employ the cluster expansion method in conjunction with density functional theory (DFT) to identify energetically stable configurations, also known as optimal colorings, of \ce{Li2B3C} and \ce{Li3B4C2} \cite{sanchez1984generalized, Fontaine1994, Zunger1994}. Notably, these derived structures deviate significantly from previously proposed configurations with same chemical stoichiometries \cite{gao2015prediction}, particularly in their atomic arrangements. Our optimal B-C layer configurations are composed of alternating B-B and B-C zigzag chains, whose electronic structures and electron-phonon coupling (EPC) properties differ from the previous calculations.

 To further investigate these properties, we calculate the electron-phonon coupling properties using the Wannier interpolation technique \cite{Giustino2007, ponce2016epw, Lee2023epw} and find that the critical temperature ($T_c$) of \ce{Li2B3C} is unexpectedly low, less than 0.03 K. However, when compressive uniaxial strain (-5\%) is applied, the electronic structure of \ce{Li2B3C} is significantly modified, dramatically increasing $T_c$ by over four orders of magnitude to approximately 37 K. This remarkable result provides valuable insights into the complex interplay between electronic structure and phonon-mediated superconductivity.

 This paper is organized as follows: In Sec. II, we optimize the lattice structures of \ce{Li_nB_{n+1}C_{n-1}} and compare them to previously reported configurations. Sec. III focuses on the analysis of electronic structures. In Sec. IV, we investigate phonon-mediated superconductivity and evaluate the deformation potential. Sec. V explores the impact of strain on both phonon-mediated superconductivity and electronic structures. Finally, we present our discussion and conclusions in Sec. VI. 
 
 For clarity, we abbreviate the previously proposed \ce{Li2B3C}/\ce{Li3B4C2} structures from Gao, Lu, and Xiang's work \cite{gao2015prediction} as \ce{Li2B3C}/\ce{Li3B4C2} (GLX), while the configurations proposed in this study are denoted as \ce{Li2B3C}/\ce{Li3B4C2} (TW).

\section{optimal coloring of L\lowercase{i}$_n$B$_{n+1}$C$_{n-1}$ ($n$=2, 3)}

\begin{table*}[ht]
\caption{\label{energy}%
The energies, space gruops and lattice parameters for \ce{Li2B3C} (TW/GLX) and \ce{Li3B4C2} (TW/GLX) after structural relaxation. The ground state energy for each formula is set as 0. Note that the lattice parameters here are for conventional cells while we use the primitive cell of \ce{Li3B4C2} (TW) in our calculations. }
\begin{ruledtabular}
\begin{tabular}{ccccccc}
  system & energy per formula (eV) & space group & $a$ (\AA) & $b$ (\AA) & $c$ (\AA) \\
 \colrule
 \ce{Li2B3C} (TW)      & 0  & $Pmma$            & 2.71 & 5.11 & 7.21    \\
  \ce{Li2B3C} (GLX) &  0.327 & $P\overline{6}m2$ & 2.84 & 2.84 & 7.13    \\
  \ce{Li3B4C2} (TW)      & 0  &$Cmcm$           & 14.93 & 2.73 & 7.17    \\
  \ce{Li3B4C2} (GLX) & 0.354  & $P\overline{6}m2$ & 4.88 & 4.88 & 3.52   \\
\end{tabular}
\end{ruledtabular}
\end{table*}

\begin{figure*}[ht]
  \centerline{\includegraphics[width=1\textwidth]{FIG1.pdf}}
  \caption{Crystal structures of \ce{Li_nB_{n+1}C_{n-1}}: (a-b) \ce{Li2B3C} and \ce{Li3B4C2} with optimal configurations from this work (TW), (c-d) \ce{Li2B3C} and \ce{Li3B4C2} proposed by Gao, Lu, and Xiang (GLX) \cite{gao2015prediction}. In (a-b), blue lines, green lines, blue dots, and red dots correspond to B-C (zigzag chain), B-B (zigzag chain), B-C (bridge), and B-B (bridge) $\sigma$-bonding WFs, respectively. Green triangles in (d) indicate three nearest-neighbouring B-B $\sigma$-bonding Wannier functions (WFs). Atom color key: light blue (Li), red (B), yellow (C). 
		\label{fig1} }
\end{figure*}

\begin{figure}[h] \centerline{\includegraphics[width=0.5\textwidth]{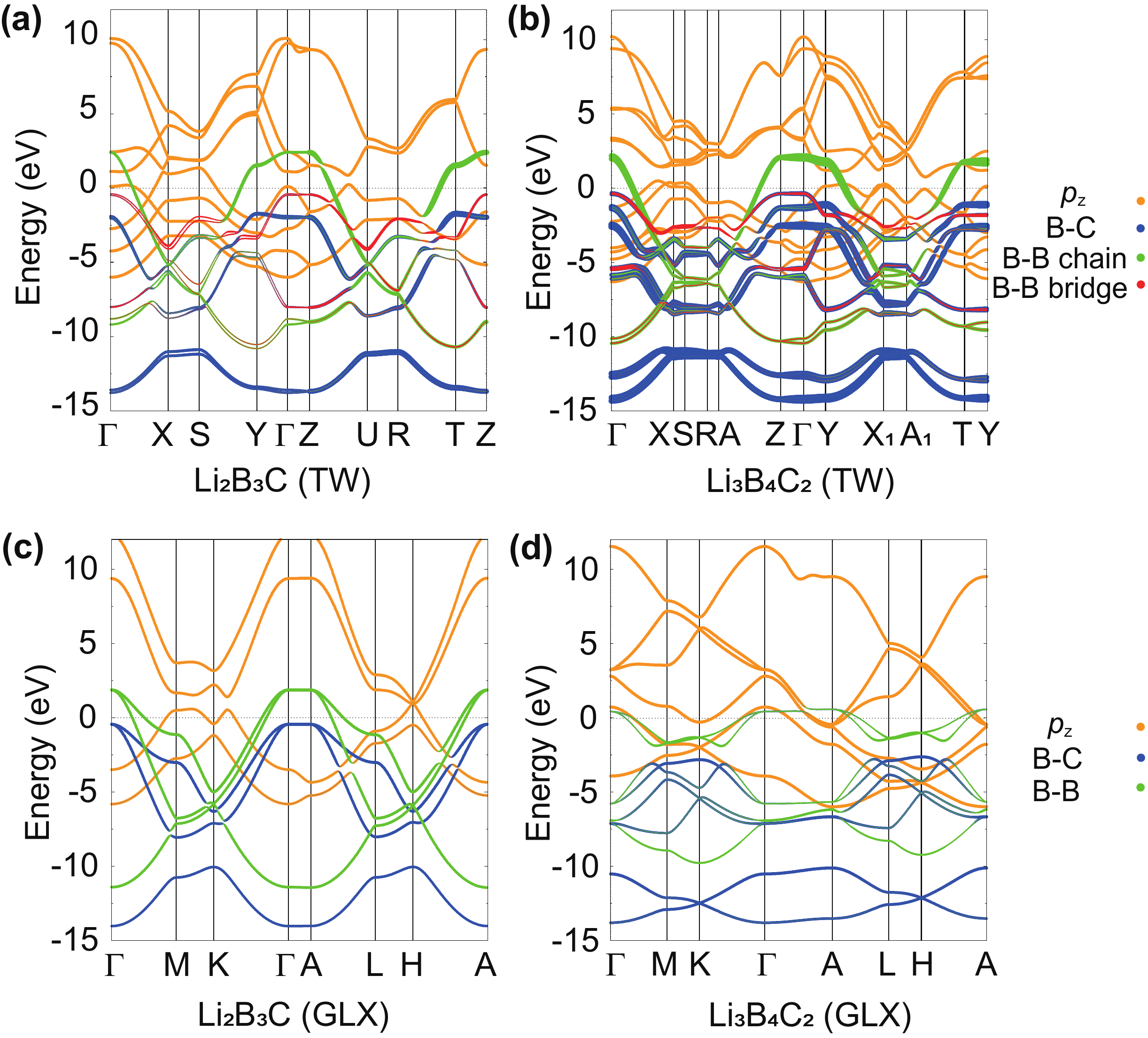}}
 \caption{Band structures for (a) \ce{Li2B3C} (TW), (b) \ce{Li3B4C2} (TW), (c) \ce{Li2B3C} (GLX), and (d) \ce{Li3B4C2} (GLX). The dot sizes are proportional to the spectral weights obtained from Wannierization. The high-symmetry \textbf{k}-points are detailed in appendix \cite{see_appendix}. 		\label{fig2} }
\end{figure}

%\begin{figure}[htb]
%\centerline{\includegraphics[width=0.5\textwidth]{FIG3.pdf}}
%\caption{\sout{Schematic band diagrams for (a) \ce{MgB2}, (b) \ce{LiBC}, (c) \ce{Li2B3C} (GLX), (d) \ce{Li3B4C2} (GLX), (e) \ce{Li2B3C}/\ce{Li3B4C2} (TW).}
%		\label{fig3} }
%\end{figure}

 To explore the optimal configuration, or "the coloring problem", of B and C atoms within \ce{B_{n+1}C_{n-1}} layer, we begin with LiBC, whose lattice structure has been well established experimentally. As depicted in appendix \cite{see_appendix}, the B and C atoms in LiBC form an ordered pattern, alternating at the vertices of a hexagonal lattice within each BC layer. However, when B atoms partially replace C atoms, the arrangement of B and C atoms within each \ce{B_{n+1}C_{n-1}} layer becomes more complex. This "coloring problem" has been largely overlooked in previous studies on \ce{Li_nB_{n+1}C_{n-1}} \cite{miao2013LiB1.1C0.9, bazhirov2014Li4B5C3, gao2015prediction, li2018LiB1+xC1-x}. For instance, earlier investigations on \ce{Li2B3C} ($n=2$) and \ce{Li3B4C2} ($n=3$) were based on the assumed colorings \cite{gao2015prediction}.

 To address this issue, we investigated the stable configurations of \ce{Li2B3C} and \ce{Li3B4C2} using the Cluster Expansion method \cite{see_appendix}. Our results (Figs. \ref{fig1}(a-b)) reveal that these colorings are markedly more stable than previously reported configurations with comparable stoichiometries and lattice structures \cite{gao2015chinese}, as detailed in Table \ref{energy}. Precisely, in the most stable configurations, B and C atoms are not uniformly distributed within each \ce{B_{n+1}C_{n-1}} layer. Instead, as highlighted in Figs. \ref{fig1}(a-b) by green and blue lines, our \ce{B_{n+1}C_{n-1}} layers can be visualized as assemblies of alternating B-B and B-C zigzag chains, as highlighted by blue and green lines in Figs. \ref{fig1}(a-b). Moreover, these zigzag chains are interconnected by what we term "bridge bonds" in this study, represented by red and blue dots in the same figures. Compared with earlier work, these unique arrangements significantly differentiate the colorings in terms of their electronic and electron-phonon coupling properties.

The crystal structures of previously predicted \ce{Li2B3C} (GLX) and \ce{Li3B4C2} (GLX) \cite{gao2015prediction} are shown in Figs. \ref{fig1}(c-d). These structures represent two principal variations within the \ce{Li_nB_{n+1}C_{n-1}} series. The first contains distinct layers of pure B atoms, mixed BC layers, and intervening Li atoms, as exemplified by \ce{Li2B3C} (GLX) in Fig. \ref{fig1}(c). Namely, \ce{Li2B3C} (GLX) can be construed as a LiBC derivative where half of the B layers are replaced by mixed BC layers. However, the elongated B-B bonds compared to B-C bonds introduce additional energy penalty, as \ce{Li2B3C} (GLX) must invest more energy to compress the pure boron layers and expand the boron-carbon layers.

Another principle variation of \ce{Li_nB_{n+1}C_{n-1}} involves uniform substitution of C atoms by B atoms in each \ce{B_{n+1}C_{n-1}} layer, such as \ce{Li3B4C2} (GLX) or our low-energy structures. In \ce{Li3B4C2} (GLX), a striking characteristic different from our optimized colorings is the absence of infinitely linked B-B zigzag chains, as illustrated in Fig. \ref{fig1}(d). Instead, the B-B bonds in \ce{Li3B4C2} (GLX) are confined to form isolated triangles that do not interconnect (highlighted as green triangles in Fig. \ref{fig1}(d)). 

%These B-B triangular arrangements are encompassed by six B-C $\sigma$-bonding Wannier functions (WFs), which necessitates a strong hybridization between B-B and B-C $\sigma$-bonding bands \cite{see_appendix}.

\section{electronic structures}

 We calculated the electronic band structures of \ce{Li_nB_{n+1}C_{n-1}} ($n=2$ and 3) with DFT method, as illustrated in Fig. \ref{fig2}. The results indicate a clear correlation between the electronic structures and the specific atomic arrangements of these materials. Based on chemical principles, the low-energy bands near the Fermi level are predominantly derived from the $sp^2$ $\sigma$- and $p_z$ $\pi$-bonding states of the B and C atoms. Hence, our Wannierization employs the $p_z$-like WFs centered on B or C atoms and $s$-like bonding WFs centered on the B-B or B-C bond centers.

Owing to the similar lattice structures and chemical stoichiometries, the band structures of \ce{Li_nB_{n+1}C_{n-1}} share some common features.  First, all $\sigma$-bonding bands exhibit highly two-dimensional, as a result of their \ce{MgB2}-like lattice. Besides, due to carbon's lower electronegativity compared to boron, the B-C $\sigma$-bonding bands lie at a lower energy level than their B-B counterparts. Consequently, as illustrated in Fig. \ref{fig2}, only the B-B $\sigma$-bonding and $\pi$ bands intersect the Fermi level in all \ce{Li_nB_{n+1}C_{n-1}}.

Then let us firstly revisit \ce{Li2B3C} (GLX) for comparison. In \ce{Li2B3C} (GLX), the B-B $\sigma$-bonding WFs and $\pi$ orbitals dominate the electronic physics near the Fermi level. This renders \ce{Li2B3C} (GLX) as an alternative version of \ce{MgB2}, albeit with increased hole doping: without considering self-doping from $\pi$ bands, one hole is doped in one B-B layer, so B-B $\sigma$-bonding WFs are nearly 2/3-filled.

In \ce{Li3B4C2} (GLX), three 120$^\circ$ rotationally equivalent B-B $\sigma$-bonding WFs strongly hybridize with six B-C $\sigma$-bonding WFs \cite{see_appendix}. As a result, both the B-B and B-C $\sigma$-bonding bands in \ce{Li3B4C2} (GLX) exhibit significant hybridization and intersect the Fermi level, influenced by the crystal field effects and boron substitution, as illustrated in Fig. \ref{fig2}(d). While it is established that one hole is doped into the nine $\sigma$-bonding bands of the \ce{B4C2^{3-}} layer (excluding contributions from self-doping through $\pi$ bands), the strong hybridization complicates the precise determination of the exact doping level.

However, compared with earlier work, an unique characteristic emerges in the electronic structures of our optimal coloring configurations, as a result of their structural characteristics. As shown in Figs. \ref{fig1}(a-b) and detailed in TABLE \ref{onsite}, the B-B bridge bonds are shorter than those within the zigzag chains. This results in stronger repulsion between the bonding and antibonding orbitals across the bridge bonds, resulting in lower on-site energies of the B-B $\sigma$-bonding WFs in the bridge bonds than those in the zigzag chains. Consequently, the B-B $\sigma$-bonding bands from the bridge bonds (along with B-C $\sigma$-bonding bands) are fully occupied, while the B-B $\sigma$-bonding bands from the zigzag chains, in addition to the $\pi$ bands, intersect the Fermi level, as depicted in Figs. \ref{fig2}(a-b). Both compounds exhibit comparable doping levels, with approximately 1/2 hole per $\sigma$-bonding WF from the B-B zigzag chains (setting aside self-doping from $\pi$ bands). This similarity is not coincidental; the structural characteristics of these compounds ensure that all bridging $\sigma$-bonding WFs are fully occupied, thereby minimizing the electronic energy. We believe that this phenomenon may be a general feature across other \ce{Li_nB_{n+1}C_{n-1}} compounds.

The electronic structures of \ce{B_{n+1}C_{n-1}} layers can be quantitatively described using WFs within the tight-binding approximation. Tables \ref{onsite} and \ref{hopping} provide the onsite energies and hopping parameters for \ce{Li2B3C} (TW) and \ce{Li3B4C2} (GLX) from our Wannierization calculations. In these \ce{B_{n+1}C_{n-1}} layers, the onsite energies of B-B $\sigma$-bonding WFs are approximately 1.8 to 2.6 eV higher than those of B-C $\sigma$-bonding WFs. Furthermore, in \ce{Li2B3C} (TW), the onsite energy of W4, which corresponds to the bridge B-B bond, is lower than that of the zigzag chain WFs (W2/W3), leading to different doping levels across B-B $\sigma$-bonding WFs. Conversely, the nearest-neighbouring hopping parameters exhibit similar magnitudes regardless of whether the atomic configurations are optimized, suggesting that variations in hopping strength do not significantly influence the electronic structures.

\begin{table}
\caption{\label{onsite}%
The on-site energies of B-B/B-C $\sigma$-bonding WFs and bond lengths of corresponding B-B/B-C $\sigma$ bonds for \ce{Li2B3C} (TW) and \ce{Li3B4C2} (GLX). The sites of WFs W1-W6/W1-W9 are illustrated in Fig. \ref{fig1}(a/d).}
\begin{ruledtabular}
\begin{tabular}{cccc}
  \ce{Li2B3C} (TW) & Bond type & Bond length (\AA) & On-site energy (eV)  \\
 \colrule
   W1 & B-C & 1.53 & -7.53  \\
   W2/3 & B-B & 1.73 & -4.60  \\
   W4 & B-B & 1.69 & -5.12  \\
   W5/6 & B-C & 1.58 & -6.93 \\
\colrule
   \ce{Li3B4C2} (GLX) & Bond type & Bond length (\AA) & On-site energy (eV) \\
 \colrule
 W1-6 & B-C & 1.59 & -6.82 \\
 W7-9 & B-B & 1.69 & -4.47 \\
\end{tabular}
\end{ruledtabular}
\end{table}

\begin{table}
\caption{\label{hopping}%
The nearest-neighbouring hopping parameters between B-B/B-C $\sigma$-bondind WFs for \ce{Li2B3C} (TW) and \ce{Li3B4C2} (GLX). }
\begin{ruledtabular}
\begin{tabular}{ccc}
  \ce{Li2B3C} (TW) &  & Hopping integral (eV) \\
 \colrule
 W5 to W6 & B-C chain to B-C chain & -2.37 \\
 W1 to W6 & B-C bridge to B-C chain & -2.35 \\
 W2 to W3 & B-B chain to B-B chain & -2.22 \\
 W4 to W5 & B-B bridge to B-C chain & -1.47 \\
 W2 to W4 & B-B chain to B-B bridge & -1.36 \\
 W1 to W2 & B-C bridge to B-B chain & -1.30 \\
\colrule
 \ce{Li3B4C2} (GLX) &  & Hopping integral (eV)\\
\colrule
W1 to W3 & B-C to B-C & -2.38 \\
W7 to W8 & B-B to B-B & -1.61 \\
W1 to W9 & B-C to B-B & -1.59 \\
W1 to W2 & B-C to B-C & -1.32 \\
\end{tabular}
\end{ruledtabular}
\end{table}

%\subsection{the analysis of the electronic structure of \ce{B_{n+1}C_{n-1}} layer}

%$T_c$ of \ce{Li2B3C} (GLX) is close to that of \ce{Li3B4C2} (GLX) \cite{gao2015prediction}.

\begin{figure*}[ht]	\centerline{\includegraphics[width=1\textwidth]{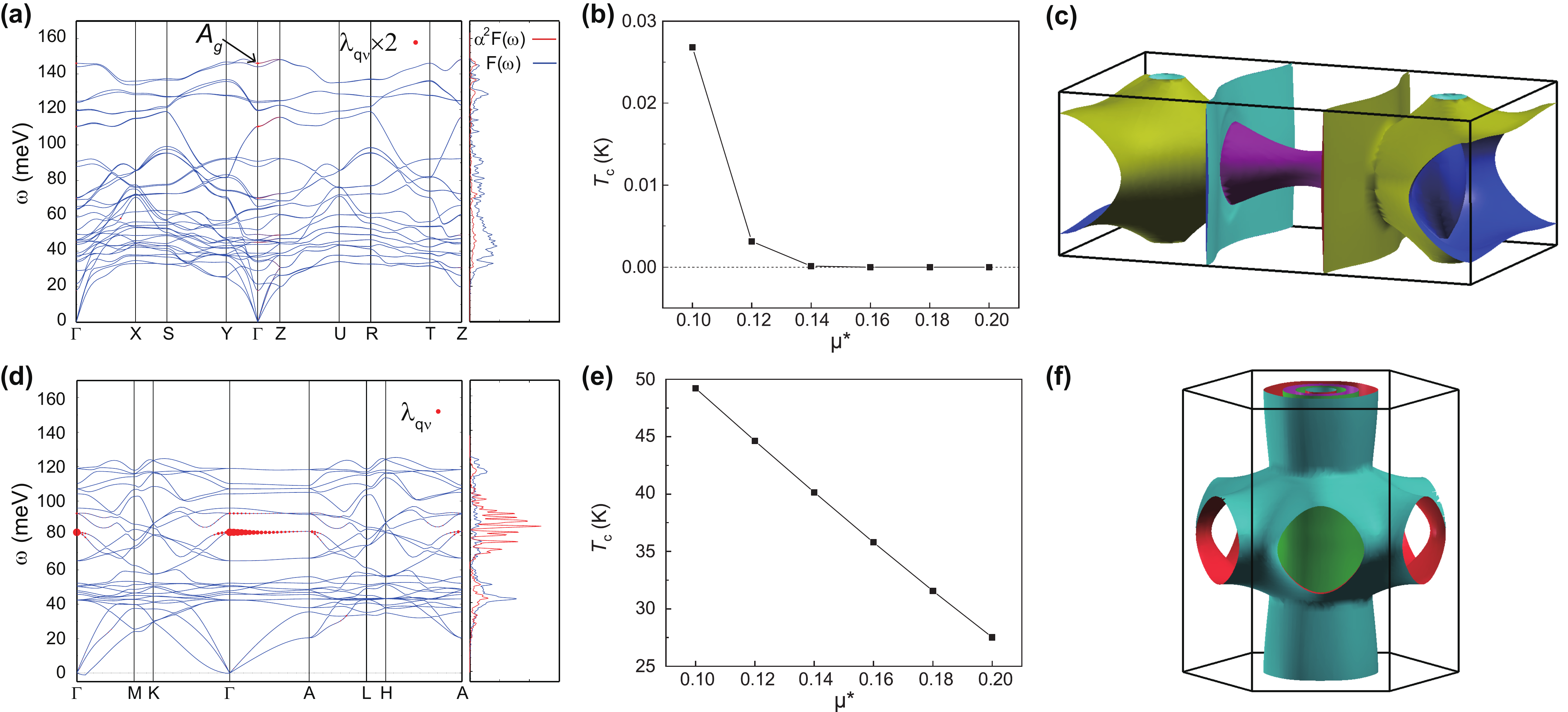}} 
 \caption{ Electron-phonon coupling constant $\lambda_{\textbf{q}\nu}$ and Fermi surface contours for (a-c) \ce{Li2B3C} (TW) and (d-f) \ce{Li3B4C2} (GLX). Panels (a) and (d) show the phonon band dispersions along with the corresponding EPC strengths $\lambda_{\textbf{q}\nu}$, where the size of the red dots indicates the magnitude of $\lambda_{\textbf{q}\nu}$. The sizes of dots are doubled in (a) for aesthetic visualization. The right subpanels display the phonon density of states (blue curves) and the Eliashberg spectral functions (red curves). Panels (b) and (e) illustrate the variation of the superconducting transition temperature $T_c$ as a function of the Coulomb pseudopotential parameter $\mu^*$. Panels (c) and (f) depict the Fermi surfaces for each compound.
		\label{fig3} }
\end{figure*}

\section{electron-phonon coupling }

\begin{table}
\caption{\label{superconductivity}%
The calculated superconducting parameters for \ce{Li2B3C} (TW) and \ce{Li3B4C2} (GLX). The $\mu^*$ here is 0.10 for $T^\mathrm{max}_c$. }
\begin{ruledtabular}
\begin{tabular}{cccc}
 Compound & $\omega_\mathrm{log}$ (meV) & $\lambda$ & $T^\mathrm{max}_c$ (K) \\
 \colrule
 \ce{Li2B3C} (TW) & 58.19 & 0.26 & 0.03  \\
 \ce{Li3B4C2} (GLX) & 59.54 & 0.96 & 49.22  \\
\end{tabular}
\end{ruledtabular}
\end{table}

 \begin{figure*}[ht]	
  \centerline{\includegraphics[width=0.9\textwidth]{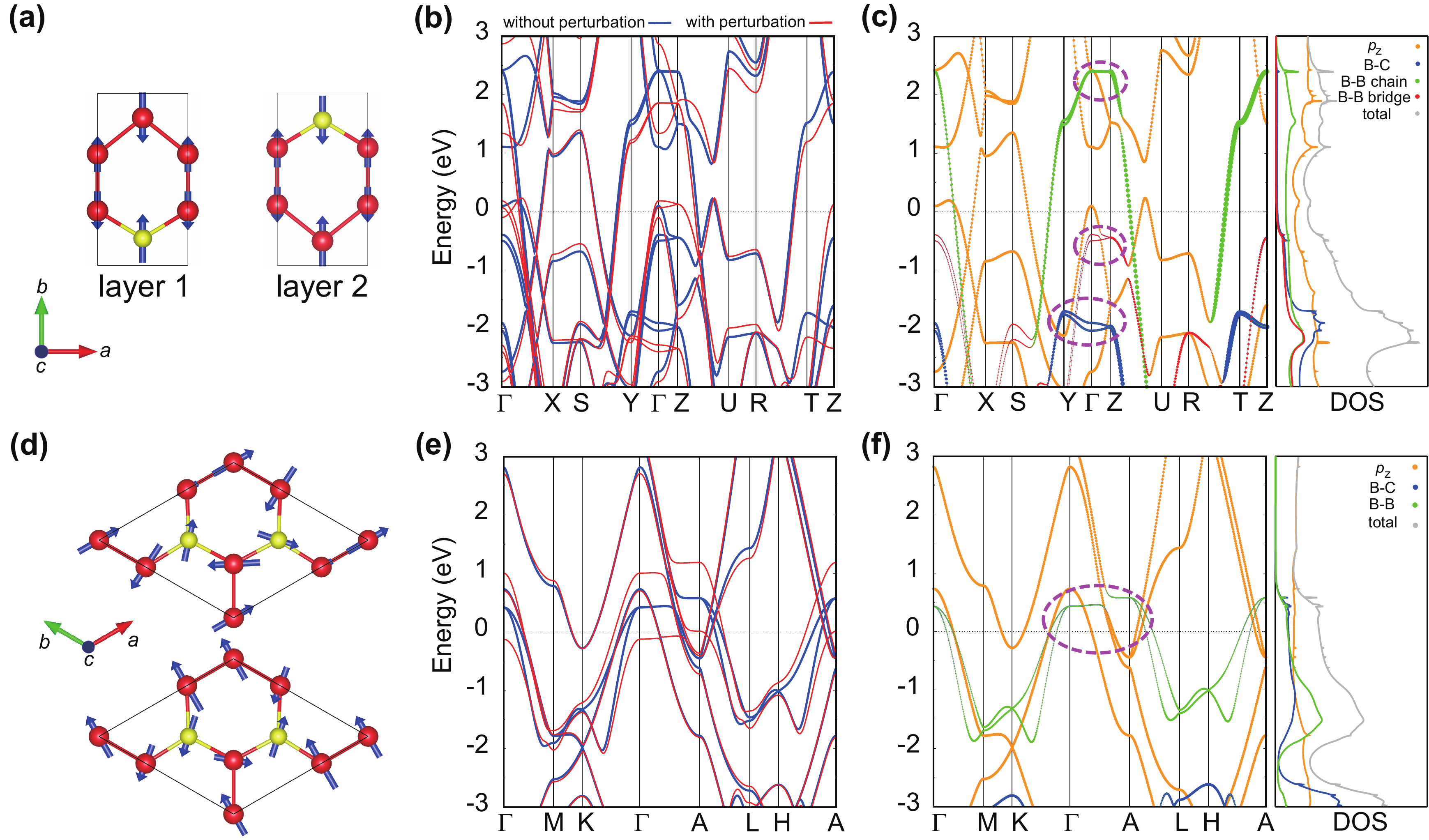}} 
  \caption{
  Effects of atomic displacements associated with specific bond-stretching phonon vibration modes on the electronic band structures for \ce{Li2B3C} (TW) and \ce{Li3B4C2} (GLX). (a) Illustration of the $A_g$ phonon mode at the $\Gamma$ point for \ce{Li2B3C} (TW), occurring at approximately 140 meV. Blue arrows indicate the directions of atomic displacements, while their lengths represent the relative amplitudes of the phonon mode. (b) Electronic band structure of \ce{Li2B3C} (TW) with a perturbation from an atomic displacement of 0.1 Å corresponding to the $A_g$ phonon mode (blue curve) compared to the band structure of the fully relaxed crystal (red curve). (c) Projected band structure and density of states (DOS) of \ce{Li2B3C} (TW), highlighting the $\sigma$-bonding bands near the $\Gamma$ point, which are sensitive to phonon vibrations, marked by dashed purple circles. (d) Illustration of the $E'$ phonon mode at the $\Gamma$ point for \ce{Li3B4C2} (GLX), occurring near 80 meV. Blue arrows indicate the directions of atomic displacements, while their lengths represent the relative amplitudes of the phonon mode. (e) Electronic band structure of \ce{Li3B4C2} (GLX) with a perturbation from an atomic displacement of 0.1 Å corresponding to the $E'$ phonon mode (blue curve) compared to the band structure of the fully relaxed crystal (red curve). (f) Projected band structure and DOS of \ce{Li3B4C2} (GLX), highlighting the $\sigma$-bonding bands near the $\Gamma$ point, which are sensitive to phonon vibrations, marked by dashed purple circles.
 \label{fig4} }
\end{figure*}

We compute the electron-phonon coupling properties and the resulting superconducting transition temperature for \ce{Li2B3C} and \ce{Li3B4C2} using the Wannier interpolation technique \cite{Giustino2007, ponce2016epw, Lee2023epw}. To provide a more detailed analysis, we focus on two specific cases: \ce{Li2B3C} (TW) and \ce{Li3B4C2} (GLX). We select \ce{Li2B3C} (TW) due to its orthogonal unit cell containing fewer atoms, which simplifies the electron-phonon coupling calculations and facilitates external strain application in subsequent studies.

We still begin by revisiting \ce{Li3B4C2} (GLX). Our findings for this compound are consistent with those reported in Ref. \cite{gao2015prediction}. As depicted in Fig. \ref{fig3}(d), the twofold degenerate bond-stretching modes, labeled as $E'$, along the $\Gamma$-A line exhibit a strong electron-phonon coupling (EPC). The corresponding coupling constant $\lambda$ and critical temperature $T_c$ are listed in Table \ref{superconductivity}. Our values are slightly lower than those in Ref. \cite{gao2015prediction}, where $\lambda = 1.114$ and $T_c \approx 55$ K. This minor deviation is attributed to the larger Gaussian smearing parameter of 0.05 Ry used in our calculations. When we reduce the smearing to 0.02 Ry, matching the value used in Ref. \cite{gao2015prediction}, our results closely align, yielding $\lambda = 1.135$ and $T_c = 58.34$ K.

 However, as shown in Fig. \ref{fig3}(a), the EPC and the corresponding Eliashberg spectral function for the optimally colored \ce{Li2B3C} are significantly weaker compared to those for \ce{Li3B4C2} (GLX). According to the McMillan-Allen-Dynes formula \cite{allen1972neutron, allen1975transition}, the superconducting transition temperature $T_c$ is determined by both the logarithmic average phonon frequency ($\omega_\mathrm{log}$) and the total EPC strength ($\lambda$) in phonon-mediated superconductors. As listed in Table \ref{superconductivity}, the $\omega_\mathrm{log}$ of \ce{Li2B3C} (TW) (58.19 meV) is comparable to that of \ce{Li3B4C2} (GLX) (59.54 meV), but the $\lambda$ value is much lower. This results in an unexpectedly low $T_c$ of approximately 0.03 K for \ce{Li2B3C} (TW), as shown in Fig. \ref{fig3}(b). Similarly, the $T_c$ of \ce{Li3B4C2} (TW) is also considerably lower than that of \ce{Li3B4C2} (GLX).

\ce{Li2B3C} (TW) and \ce{Li3B4C2} (GLX) also exhibit notable differences in their Fermi surface structures, as shown in Figs. \ref{fig3}(c) and \ref{fig3}(f). For both compounds, the Fermi surfaces corresponding to the $\sigma$-bonding bands are relatively flat along the $k_z$ direction due to the weak interlayer coupling of the $\sigma$-bonding WFs. However, the $\sigma$-bonding Fermi surfaces of \ce{Li2B3C} (TW) are strongly 2D-anisotropic, compared to those of \ce{Li3B4C2} (GLX). This is consistent with the greater dispersion of the $\sigma$-bonding bands along the $\Gamma$-X direction in \ce{Li2B3C} (TW).

EPC is governed by the deformation potential $D_{\textbf{q}\nu}$, as described by the equation \cite{khan1984deformation, mgb2_pickett, lee2004superconductivity, lee2005crystal, tan2021dp}:
\begin{equation}\label{dp-epc} 
\lambda_{\textbf{q}\nu}=N(0)\frac{|D_{\textbf{q}\nu}|^2}{M_{\textbf{q}\nu}\omega_{\textbf{q}\nu}^2}, 
\end{equation}
where $D_{\textbf{q}\nu}$ denotes the shift in band energy resulting from a perturbation of the phonon mode $\textbf{q}\nu$ near the Fermi surface, $M_{\textbf{q}\nu}$ is the averaged atomic mass, and $\omega_{\textbf{q}\nu}$ is the phonon frequency. The strong covalent nature of $\sigma$ bands results in a large deformation potential, giving a direct physical picture of the phonon-mediated high-temperature superconductivity of MgB$_2$ \cite{mgb2_pickett} or \ce{Li3B4C2} (GLX) \cite{gao2015prediction}. 

To understand the significant suppression of EPC in the optimal coloring compounds, we examine how the band structures are affected by perturbations in the bond-stretching phonon modes. Fig. \ref{fig4}(b/e) presents the band structure of \ce{Li2B3C} (TW)/\ce{Li3B4C2} (GLX) before and after introducing the perturbation, respectively. For \ce{Li2B3C} (TW), we selected the $A_g$ phonon mode (Fig. \ref{fig4}(a)) as a representative bond-stretching phonon mode due to its strongest electron-phonon coupling (EPC) at the $\Gamma$ point. For \ce{Li3B4C2} (GLX), we chose the $E'$ mode shown in Fig. \ref{fig4}(d), same as in Ref. \cite{gao2015prediction}.

A common observation for both \ce{Li2B3C} (TW) and \ce{Li3B4C2} (GLX) is the significant shift at the top of the $\sigma$-bonding bands upon perturbation, as illustrated in Figs. \ref{fig4}(b,e) and highlighted by the dashed purple circles in Figs. \ref{fig4}(c,f). This finding is consistent with previous work on \ce{Li3B4C2} (GLX) \cite{gao2015prediction}, where the doubly degenerate $\sigma$-bonding bands split by approximately 1.1 eV along the $\Gamma$-A direction under the influence of the $E'$ bond-stretching phonon mode, suggesting a strong EPC \cite{gao2015prediction}. This behavior arises because the Fermi level in \ce{Li3B4C2} (GLX) lies near the top of the $\sigma$-bonding bands.

In contrast, for \ce{Li2B3C} (TW), the top of the $\sigma$-bonding bands is either positioned below or well above the Fermi level, indicating that the states at the Fermi level predominantly arise from more dispersive states, which are much less sensitive to lattice vibrations. From a chemical bonding perspective \cite{hoffmann1991COOP, dronskowski1993COHP}, these dispersive states are primarily contributed by nonbonding electrons. Consequently, when the perturbation is introduced, the bonding and antibonding extrema of bands exhibit significant changes, while the nonbonding dispersive states remain nearly unaffected. Furthermore, the electronic nature of such scenario is that the Fermi level is positioned near 3/4-filling in \ce{Li2B3C} (TW), as we analyzed in last section. Hence, small deformation potential and weak EPC might be general across other \ce{Li_nB_{n+1}C_{n-1}} compounds.

The above analysis indicates that achieving a large EPC requires not only the presence of $\sigma$-bonding character but also a significant deformation potential. Therefore, for the emergence of phonon-mediated high-$T_c$ superconductivity, the states at the Fermi level must exhibit large deformation potentials, which are typically associated with strong covalent bonding or antibonding states.
 
 \section{Effect of uniaxial strain}

\begin{figure}[tb]
	 \centerline{\includegraphics[width=0.5\textwidth]{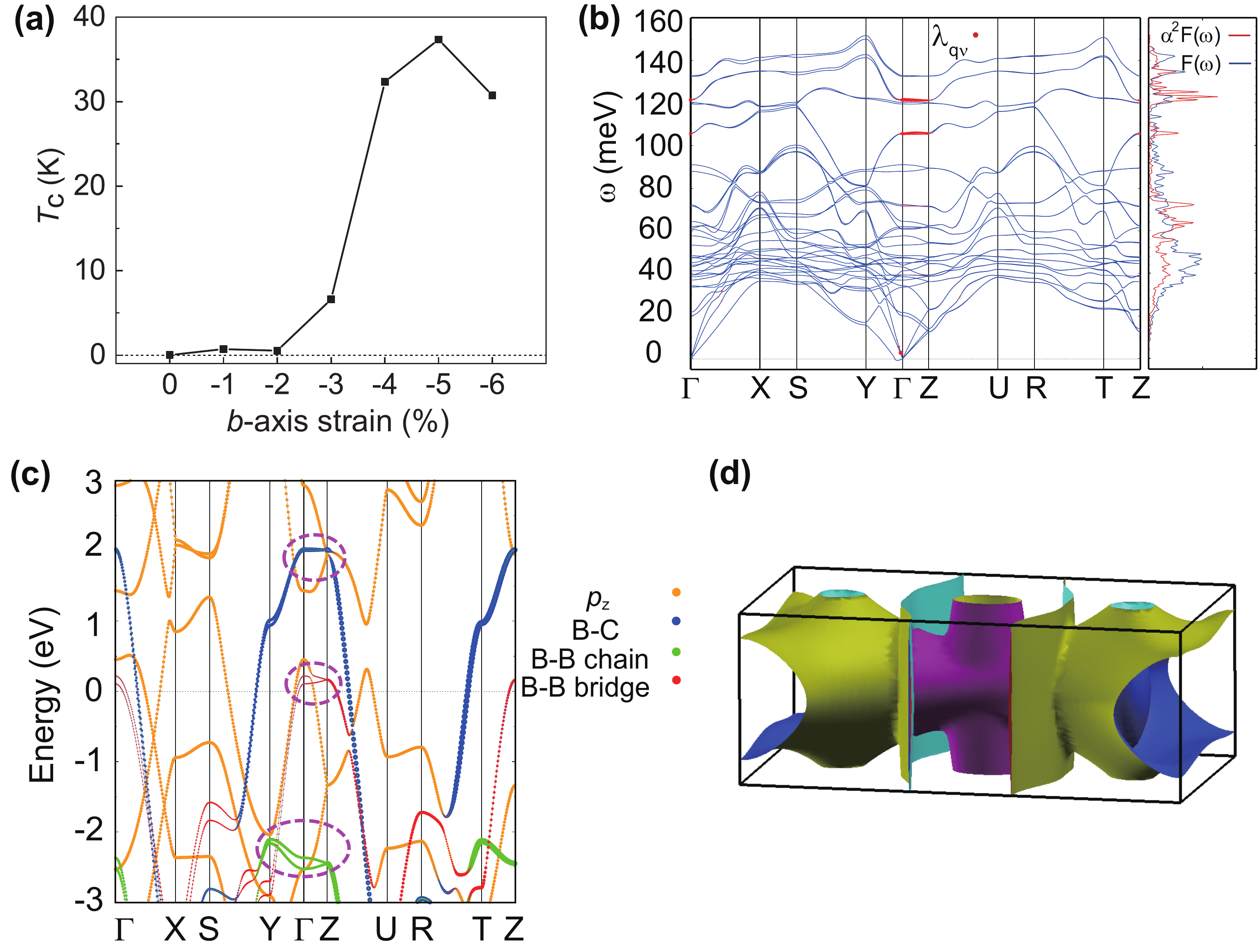}}
 \caption{Impact of compressive $b$-axis strain on the superconducting transition temperature and electronic structure of \ce{Li2B3C} (TW). (a) $T_c$ as a function of compressive $b$-axis strain. (b) Phonon dispersion under $-4\%$ compressive $b$-axis strain, where the red dot size is proportional to the electron-phonon coupling strength, $\lambda_{\textbf{q}\nu}$. The right panel displays the phonon density of states (blue line) and the Eliashberg spectral function (red line). (c) Projected band structure under $-4\%$ compressive $b$-axis strain, with dot sizes representing the spectral weights. The top of the $\sigma$-bonding bands near the $\Gamma$ point, sensitive to phonon vibrations, are indicated by dashed purple circles. (d) Fermi surfaces under $-4\%$ compressive $b$-axis strain.
 \label{strain} }
\end{figure}

\begin{table}
\caption{\label{onsite-strain}%
On-site energies of B-B or B-C $\sigma$-bonding WFs with respect to the Fermi level and the bond lengths of B-B or B-C $\sigma$ bonds for \ce{Li2B3C} (TW) with $-4\%$ compressive $b$-axis strain. The sites of WFs W1-W6 are illustrated in Fig. \ref{fig1}(a). }
\begin{ruledtabular}
\begin{tabular}{cccc}
  WF & Bond type & Bond length (\AA) & On-site energy (eV)  \\
 \colrule
   W1 & B-C & 1.51 & -7.38  \\
   W2/3 & B-B & 1.70 & -4.61  \\
   W4 & B-B & 1.64 & -4.95  \\
   W5/6 & B-C & 1.58 & -6.95 \\
\end{tabular}
\end{ruledtabular}
\end{table}

\begin{table}[b]
\caption{\label{hopping-strain}%
Nearest-neighbouring hopping parameters between B-B/B-C $\sigma$-bondind WFs for \ce{Li2B3C} (TW) with a $-4\%$ compressive $b$-axis strain. }
\begin{ruledtabular}
\begin{tabular}{ccc}
\ce{Li2B3C} (TW, $-4\%$)  &   & Hopping integral (eV) \\
 \colrule
 W5 to W6 & B-C chain to B-C chain & -2.25 \\
 W1 to W6 & B-C brdige to B-C chain & -2.44 \\
 W2 to W3 & B-B chain to B-B chain & -2.04 \\
 W4 to W5 & B-B bridge to B-C chain & -1.58 \\
 W2 to W4 & B-B chain to B-B bridge & -1.46 \\
 W1 to W2 & B-C bridge to B-B chain & -1.39 \\
\end{tabular}
\end{ruledtabular}
\end{table}

 The above analysis suggests that EPC is dramatically suppressed by lattice deformations in the optimal configuration of \ce{Li_nB_{n+1}C_{n-1}}, compared to the \ce{Li_nB_{n+1}C_{n-1}} (GLX) phase. However, since the $\sigma$-bonding bands already emerge at the Fermi level, there remains potential to significantly strengthen EPC by tuning the electronic structure to harness the strong covalent bonds in these systems, for instance, through the application of strain. In the cases of both \ce{Li2B3C} (TW) and \ce{Li3B4C2} (TW), we observe that EPC can be significantly enhanced by applying moderate compressive strain along the $b$-axis.

 Fig. \ref{strain}(a) illustrates how $T_c$ varies with applied strain for \ce{Li2B3C} (TW), calculated using the McMillan-Allen-Dynes formula \cite{allen1972neutron, allen1975transition} with an effective Coulomb potential parameter of $\mu^* = 0.10$. Under -3\% $b$-axis strain, $T_c$ rises to 6.61 K and continues to increase with further strain, forming a dome-like dependence and peaking at 37 K under -5\% $b$-axis strain.

Similar to the behavior observed in \ce{MgB2} and \ce{Li3B4C2} (GLX), the EPC in \ce{Li2B3C} (TW) is significantly enhanced for bond-stretching phonon modes along the $\Gamma$-Z direction when subjected to compressive strain along the $b$-axis. For example, this enhancement is evident in the changes to both the EPC and electronic structure of \ce{Li2B3C} (TW) under $-4\%$ $b$-axis strain, as shown in Figs. \ref{strain}(b-c). The dramatic increase in EPC is associated with significant changes in the electronic structure. As illustrated in Figs. \ref{strain}(c-d), the bridge B-B $\sigma$-bonding bands cross the Fermi level, giving rise to elliptical cylindrical Fermi surfaces near the $\Gamma$ point. Notably, the top of these bridge B-B $\sigma$-bonding bands now are above the Fermi level, as highlighted by the dashed purple circles in Fig. \ref{strain}(c). This alignment generates a large deformation potential, leading to a substantial boost in EPC.

The application of compressive strain primarily alters the bond angles and slightly reduces the bond lengths along the $b$-axis in \ce{Li2B3C} (TW). Concurrently, the lattice constants $a$ and $c$ experience minor elongation. These structural modifications lead to an increase in the hopping parameters from the bridge sites to the zigzag chain sites, while the hopping parameters within the zigzag chains decrease, as shown in Table \ref{hopping-strain}. For instance, the hopping integral between W4 and W2 changes from -1.36 eV to -1.46 eV. This adjustment enhances the hybridization between the bridge B-B $\sigma$-bonding bands and the zigzag chain B-B $\sigma$-bonding bands. Additionally, the compressive strain along the $b$-axis raises the on-site energies of the $\sigma$-bonding WFs at the bridge sites for both B-B and B-C bonds, as detailed in Table \ref{onsite-strain}.

 Analyzing Eq. \eqref{dp-epc}, it is important to emphasize the critical role of DOS at the Fermi level in determining EPC \cite{sano2016VHS, gai2022B3C3, geng2023B3C3}. For instance, despite a strong deformation potential, a low DOS can yield weak EPC, as seen in boron-doped diamond \cite{lee2004superconductivity}. This phenomenon explains why the superconducting transition temperature ($T_c$) remains low under -3\% compressive $b$-axis strain. Slightly below the  -3\% compressive $b$-axis strain, the bridge B-B $\sigma$-bonding bands just intersect the Fermi level, triggering a Lifshitz transition. However, these bands contribute negligibly to DOS at the Fermi level, leading to weak EPC in \ce{Li2B3C} (TW) \cite{see_appendix}. As the strain is further increased, more hole are doped in the bridge B-B $\sigma$-bonding WFs, enlarging the Fermi surfaces (mainly comprised of bridge B-B $\sigma$-bonding WFs) and boosting DOS, even though the deformation potential decreases \cite{see_appendix}. The interplay between these two opposing factors results in the characteristic dome-shaped strain-$T_c$ relationship.

\section{Discussion and Conclusion}

  In summary, we identify the optimal configurations of \ce{Li2B3C} and \ce{Li3B4C2} using the cluster expansion method, demonstrating that these configurations are thermodynamically more stable than previously reported structures. In these optimized configurations, both B-B and B-C zigzag chains emerge within each \ce{B_{n+1}C_{n-1}} plane, resulting in unique electronic structures that differ qualitatively from those reported in earlier studies \cite{miao2013LiB1.1C0.9, bazhirov2014Li4B5C3, gao2015prediction, li2018LiB1+xC1-x}. Notably, in \ce{Li_nB_{n+1}C_{n-1}} (TW) ($n=2,3$), only the $\sigma$-bonding bands associated with the B-B zigzag chains cross the Fermi level, exhibiting weak hybridization with the $\sigma$-bonding bands in the B-C zigzag chains. Additionally, the $\sigma$-bonding bands in the B-B zigzag chains are approximately 3/4-filled and display significant dispersion along their corresponding directions in reciprocal space. In contrast, other $\sigma$-bonding bands remain fully occupied, including the bridge B-B and all B-C $\sigma$-bonding bands.

  Our calculations of EPC and deformation potential for \ce{Li3B4C2} (GLX) and \ce{Li2B3C} (TW) reveal that both the EPC and deformation potential are significantly weaker in \ce{Li2B3C} (TW) than in \ce{Li3B4C2} (GLX). As a result, the superconducting transition temperature of \ce{Li2B3C} (TW) is less than 0.03 K, dramatically lower than the predicted $T_c$ of 50 K for \ce{Li3B4C2} (GLX). Nonetheless, since the $\sigma$-bonding bands in \ce{Li2B3C} (TW) are already present above the Fermi level, there remains significant potential to transform \ce{Li2B3C} (TW) into a phonon-mediated high-$T_c$ superconductor by applying a uniaxial strain. Our strain-dependent calculations show that applying a -5\% compressive strain along the $b$-axis in \ce{Li2B3C} (TW) can markedly alter the electronic band structure around the Fermi level and significantly enhance the deformation potential on the Fermi surface, thereby raising $T_c$ to approximately 37 K. Notably, deformation potential can be measured by modern experimental techniques nowadays \cite{ZXShen2017elph_dp,huang2023ultrafast,sobota2021angle,boschini2024time}. 
  
 Our work demonstrates that not only the chemical stoichiometry but also the optimal coloring can significantly influence the superconducting pairing interactions. Therefore, it is meaningful to investigate the true coloring of those doped superconductors \cite{bhaumik2017high, sun2023, ding2023critical}. Moreover, our findings highlight the critical role of the deformation potential on electrons at the Fermi level, which effectively measures the EPC and is closely linked to the nature of chemical bonding.

 Furthermore, our work underscores the need to establish a robust criterion for quantifying the contribution of $\sigma$-bonding electrons at the Fermi level. Such a criterion would serve as a key metric for evaluating the strength of electronic interactions and provide a reliable indicator for identifying materials with high-$T_c$ superconducting potential. Implementing this metric could streamline the discovery of novel high-$T_c$ superconductors through high-throughput searches, enabling more efficient exploration of candidate materials. By accurately assessing the role of $\sigma$-bonding electrons and deformation potentials, it is possible to narrow the range of potential compounds and prioritize those with electronic structures favorable for high-temperature superconductivity. This systematic approach is essential not only for optimizing the search process and minimizing experimental efforts but also for advancing the understanding of the fundamental mechanisms that govern high-$T_c$ superconductivity.

{\it Acknowledgement:}
We thank Xi Xu and Yue-chao Wang for useful discussions. The work is supported by the National Science Foundation of China (Grant No: 12488201, 11888101), the Ministry of Science and Technology (Grant No: 2021ZD0301800).
\clearpage
\bibliography{main}

\begin{thebibliography}{76}%
\makeatletter
\providecommand \@ifxundefined [1]{%
 \@ifx{#1\undefined}
}%
\providecommand \@ifnum [1]{%
 \ifnum #1\expandafter \@firstoftwo
 \else \expandafter \@secondoftwo
 \fi
}%
\providecommand \@ifx [1]{%
 \ifx #1\expandafter \@firstoftwo
 \else \expandafter \@secondoftwo
 \fi
}%
\providecommand \natexlab [1]{#1}%
\providecommand \enquote  [1]{``#1''}%
\providecommand \bibnamefont  [1]{#1}%
\providecommand \bibfnamefont [1]{#1}%
\providecommand \citenamefont [1]{#1}%
\providecommand \href@noop [0]{\@secondoftwo}%
\providecommand \href [0]{\begingroup \@sanitize@url \@href}%
\providecommand \@href[1]{\@@startlink{#1}\@@href}%
\providecommand \@@href[1]{\endgroup#1\@@endlink}%
\providecommand \@sanitize@url [0]{\catcode `\\12\catcode `\$12\catcode
  `\&12\catcode `\#12\catcode `\^12\catcode `\_12\catcode `\%12\relax}%
\providecommand \@@startlink[1]{}%
\providecommand \@@endlink[0]{}%
\providecommand \url  [0]{\begingroup\@sanitize@url \@url }%
\providecommand \@url [1]{\endgroup\@href {#1}{\urlprefix }}%
\providecommand \urlprefix  [0]{URL }%
\providecommand \Eprint [0]{\href }%
\providecommand \doibase [0]{http://dx.doi.org/}%
\providecommand \selectlanguage [0]{\@gobble}%
\providecommand \bibinfo  [0]{\@secondoftwo}%
\providecommand \bibfield  [0]{\@secondoftwo}%
\providecommand \translation [1]{[#1]}%
\providecommand \BibitemOpen [0]{}%
\providecommand \bibitemStop [0]{}%
\providecommand \bibitemNoStop [0]{.\EOS\space}%
\providecommand \EOS [0]{\spacefactor3000\relax}%
\providecommand \BibitemShut  [1]{\csname bibitem#1\endcsname}%
\let\auto@bib@innerbib\@empty
%</preamble>
\bibitem [{\citenamefont {Bednorz}\ and\ \citenamefont
  {M{\"u}ller}(1986)}]{bednorz1986possible}%
  \BibitemOpen
  \bibfield  {author} {\bibinfo {author} {\bibfnamefont {J.~G.}\ \bibnamefont
  {Bednorz}}\ and\ \bibinfo {author} {\bibfnamefont {K.~A.}\ \bibnamefont
  {M{\"u}ller}},\ }\href@noop {} {\bibfield  {journal} {\bibinfo  {journal}
  {Zeitschrift f{\"u}r Physik B Condensed Matter}\ }\textbf {\bibinfo {volume}
  {64}},\ \bibinfo {pages} {189} (\bibinfo {year} {1986})}\BibitemShut
  {NoStop}%
\bibitem [{\citenamefont {Nagamatsu}\ \emph {et~al.}(2001)\citenamefont
  {Nagamatsu}, \citenamefont {Nakagawa}, \citenamefont {Muranaka},
  \citenamefont {Zenitani},\ and\ \citenamefont
  {Akimitsu}}]{nagamatsu2001superconductivity}%
  \BibitemOpen
  \bibfield  {author} {\bibinfo {author} {\bibfnamefont {J.}~\bibnamefont
  {Nagamatsu}}, \bibinfo {author} {\bibfnamefont {N.}~\bibnamefont {Nakagawa}},
  \bibinfo {author} {\bibfnamefont {T.}~\bibnamefont {Muranaka}}, \bibinfo
  {author} {\bibfnamefont {Y.}~\bibnamefont {Zenitani}}, \ and\ \bibinfo
  {author} {\bibfnamefont {J.}~\bibnamefont {Akimitsu}},\ }\href@noop {}
  {\bibfield  {journal} {\bibinfo  {journal} {Nature}\ }\textbf {\bibinfo
  {volume} {410}},\ \bibinfo {pages} {63} (\bibinfo {year} {2001})}\BibitemShut
  {NoStop}%
\bibitem [{\citenamefont {Drozdov}\ \emph {et~al.}(2015)\citenamefont
  {Drozdov}, \citenamefont {Eremets}, \citenamefont {Troyan}, \citenamefont
  {Ksenofontov},\ and\ \citenamefont {Shylin}}]{drozdov2015conventional}%
  \BibitemOpen
  \bibfield  {author} {\bibinfo {author} {\bibfnamefont {A.}~\bibnamefont
  {Drozdov}}, \bibinfo {author} {\bibfnamefont {M.}~\bibnamefont {Eremets}},
  \bibinfo {author} {\bibfnamefont {I.}~\bibnamefont {Troyan}}, \bibinfo
  {author} {\bibfnamefont {V.}~\bibnamefont {Ksenofontov}}, \ and\ \bibinfo
  {author} {\bibfnamefont {S.~I.}\ \bibnamefont {Shylin}},\ }\href@noop {}
  {\bibfield  {journal} {\bibinfo  {journal} {Nature}\ }\textbf {\bibinfo
  {volume} {525}},\ \bibinfo {pages} {73} (\bibinfo {year} {2015})}\BibitemShut
  {NoStop}%
\bibitem [{\citenamefont {Gao}\ \emph {et~al.}(2015{\natexlab{a}})\citenamefont
  {Gao}, \citenamefont {Lu},\ and\ \citenamefont {Xiang}}]{gao2015chinese}%
  \BibitemOpen
  \bibfield  {author} {\bibinfo {author} {\bibfnamefont {M.}~\bibnamefont
  {Gao}}, \bibinfo {author} {\bibfnamefont {Z.-Y.}\ \bibnamefont {Lu}}, \ and\
  \bibinfo {author} {\bibfnamefont {T.}~\bibnamefont {Xiang}},\ }\href@noop {}
  {\bibfield  {journal} {\bibinfo  {journal} {PHYSICS (in Chinese)}\ }\textbf
  {\bibinfo {volume} {44}},\ \bibinfo {pages} {421} (\bibinfo {year}
  {2015}{\natexlab{a}})}\BibitemShut {NoStop}%
\bibitem [{\citenamefont {An}\ and\ \citenamefont
  {Pickett}(2001)}]{mgb2_pickett}%
  \BibitemOpen
  \bibfield  {author} {\bibinfo {author} {\bibfnamefont {J.~M.}\ \bibnamefont
  {An}}\ and\ \bibinfo {author} {\bibfnamefont {W.~E.}\ \bibnamefont
  {Pickett}},\ }\href {\doibase 10.1103/PhysRevLett.86.4366} {\bibfield
  {journal} {\bibinfo  {journal} {Phys. Rev. Lett.}\ }\textbf {\bibinfo
  {volume} {86}},\ \bibinfo {pages} {4366} (\bibinfo {year}
  {2001})}\BibitemShut {NoStop}%
\bibitem [{\citenamefont {Gao}\ \emph {et~al.}(2015{\natexlab{b}})\citenamefont
  {Gao}, \citenamefont {Lu},\ and\ \citenamefont {Xiang}}]{gao2015prediction}%
  \BibitemOpen
  \bibfield  {author} {\bibinfo {author} {\bibfnamefont {M.}~\bibnamefont
  {Gao}}, \bibinfo {author} {\bibfnamefont {Z.-Y.}\ \bibnamefont {Lu}}, \ and\
  \bibinfo {author} {\bibfnamefont {T.}~\bibnamefont {Xiang}},\ }\href@noop {}
  {\bibfield  {journal} {\bibinfo  {journal} {Phys. Rev. B}\ }\textbf {\bibinfo
  {volume} {91}},\ \bibinfo {pages} {045132} (\bibinfo {year}
  {2015}{\natexlab{b}})}\BibitemShut {NoStop}%
\bibitem [{\citenamefont {Deng}\ \emph {et~al.}(2003)\citenamefont {Deng},
  \citenamefont {Simon},\ and\ \citenamefont {K{\"o}hler}}]{deng2003mgb2flat}%
  \BibitemOpen
  \bibfield  {author} {\bibinfo {author} {\bibfnamefont {S.}~\bibnamefont
  {Deng}}, \bibinfo {author} {\bibfnamefont {A.}~\bibnamefont {Simon}}, \ and\
  \bibinfo {author} {\bibfnamefont {J.}~\bibnamefont {K{\"o}hler}},\
  }\href@noop {} {\bibfield  {journal} {\bibinfo  {journal} {J. Supercond.}\
  }\textbf {\bibinfo {volume} {16}},\ \bibinfo {pages} {477} (\bibinfo {year}
  {2003})}\BibitemShut {NoStop}%
\bibitem [{\citenamefont {Deng}\ \emph {et~al.}(2005)\citenamefont {Deng},
  \citenamefont {Simon},\ and\ \citenamefont {K{\"o}hler}}]{deng2005flat}%
  \BibitemOpen
  \bibfield  {author} {\bibinfo {author} {\bibfnamefont {S.}~\bibnamefont
  {Deng}}, \bibinfo {author} {\bibfnamefont {A.}~\bibnamefont {Simon}}, \ and\
  \bibinfo {author} {\bibfnamefont {J.}~\bibnamefont {K{\"o}hler}},\
  }\href@noop {} {\bibfield  {journal} {\bibinfo  {journal} {Inter. J. Mod.
  Phys. B}\ }\textbf {\bibinfo {volume} {19}},\ \bibinfo {pages} {29} (\bibinfo
  {year} {2005})}\BibitemShut {NoStop}%
\bibitem [{\citenamefont {Lee}\ and\ \citenamefont
  {Pickett}(2004)}]{lee2004superconductivity}%
  \BibitemOpen
  \bibfield  {author} {\bibinfo {author} {\bibfnamefont {K.-W.}\ \bibnamefont
  {Lee}}\ and\ \bibinfo {author} {\bibfnamefont {W.~E.}\ \bibnamefont
  {Pickett}},\ }\href@noop {} {\bibfield  {journal} {\bibinfo  {journal} {Phys.
  Rev. Lett.}\ }\textbf {\bibinfo {volume} {93}},\ \bibinfo {pages} {237003}
  (\bibinfo {year} {2004})}\BibitemShut {NoStop}%
\bibitem [{\citenamefont {Peng}\ \emph {et~al.}(2017)\citenamefont {Peng},
  \citenamefont {Sun}, \citenamefont {Pickard}, \citenamefont {Needs},
  \citenamefont {Wu},\ and\ \citenamefont {Ma}}]{peng2017hydrogen}%
  \BibitemOpen
  \bibfield  {author} {\bibinfo {author} {\bibfnamefont {F.}~\bibnamefont
  {Peng}}, \bibinfo {author} {\bibfnamefont {Y.}~\bibnamefont {Sun}}, \bibinfo
  {author} {\bibfnamefont {C.~J.}\ \bibnamefont {Pickard}}, \bibinfo {author}
  {\bibfnamefont {R.~J.}\ \bibnamefont {Needs}}, \bibinfo {author}
  {\bibfnamefont {Q.}~\bibnamefont {Wu}}, \ and\ \bibinfo {author}
  {\bibfnamefont {Y.}~\bibnamefont {Ma}},\ }\href@noop {} {\bibfield  {journal}
  {\bibinfo  {journal} {Phys. Rev. Lett.}\ }\textbf {\bibinfo {volume} {119}},\
  \bibinfo {pages} {107001} (\bibinfo {year} {2017})}\BibitemShut {NoStop}%
\bibitem [{\citenamefont {Gu}\ \emph {et~al.}(2021)\citenamefont {Gu},
  \citenamefont {Wu}, \citenamefont {Jiang},\ and\ \citenamefont
  {Hu}}]{gu2021bacus2}%
  \BibitemOpen
  \bibfield  {author} {\bibinfo {author} {\bibfnamefont {Y.}~\bibnamefont
  {Gu}}, \bibinfo {author} {\bibfnamefont {X.}~\bibnamefont {Wu}}, \bibinfo
  {author} {\bibfnamefont {K.}~\bibnamefont {Jiang}}, \ and\ \bibinfo {author}
  {\bibfnamefont {J.}~\bibnamefont {Hu}},\ }\href@noop {} {\bibfield  {journal}
  {\bibinfo  {journal} {Chin. Phys. Lett.}\ }\textbf {\bibinfo {volume} {38}},\
  \bibinfo {pages} {017501} (\bibinfo {year} {2021})}\BibitemShut {NoStop}%
\bibitem [{\citenamefont {Geng}\ \emph {et~al.}(2023)\citenamefont {Geng},
  \citenamefont {Hilleke}, \citenamefont {Zhu}, \citenamefont {Wang},
  \citenamefont {Strobel},\ and\ \citenamefont {Zurek}}]{geng2023B3C3}%
  \BibitemOpen
  \bibfield  {author} {\bibinfo {author} {\bibfnamefont {N.}~\bibnamefont
  {Geng}}, \bibinfo {author} {\bibfnamefont {K.~P.}\ \bibnamefont {Hilleke}},
  \bibinfo {author} {\bibfnamefont {L.}~\bibnamefont {Zhu}}, \bibinfo {author}
  {\bibfnamefont {X.}~\bibnamefont {Wang}}, \bibinfo {author} {\bibfnamefont
  {T.~A.}\ \bibnamefont {Strobel}}, \ and\ \bibinfo {author} {\bibfnamefont
  {E.}~\bibnamefont {Zurek}},\ }\href@noop {} {\bibfield  {journal} {\bibinfo
  {journal} {J. Am. Chem. Soc.}\ } (\bibinfo {year} {2023})}\BibitemShut
  {NoStop}%
\bibitem [{\citenamefont {Geng}\ \emph {et~al.}(2024)\citenamefont {Geng},
  \citenamefont {Hilleke}, \citenamefont {Belli}, \citenamefont {Das},\ and\
  \citenamefont {Zurek}}]{geng2024superconductivity}%
  \BibitemOpen
  \bibfield  {author} {\bibinfo {author} {\bibfnamefont {N.}~\bibnamefont
  {Geng}}, \bibinfo {author} {\bibfnamefont {K.~P.}\ \bibnamefont {Hilleke}},
  \bibinfo {author} {\bibfnamefont {F.}~\bibnamefont {Belli}}, \bibinfo
  {author} {\bibfnamefont {P.~K.}\ \bibnamefont {Das}}, \ and\ \bibinfo
  {author} {\bibfnamefont {E.}~\bibnamefont {Zurek}},\ }\href@noop {}
  {\bibfield  {journal} {\bibinfo  {journal} {Mater. Today Phys.}\ }\textbf
  {\bibinfo {volume} {44}},\ \bibinfo {pages} {101443} (\bibinfo {year}
  {2024})}\BibitemShut {NoStop}%
\bibitem [{\citenamefont {Zhang}\ \emph {et~al.}(2023)\citenamefont {Zhang},
  \citenamefont {Zhan}, \citenamefont {Gao}, \citenamefont {Liu}, \citenamefont
  {Ren}, \citenamefont {Lu},\ and\ \citenamefont {Xiang}}]{ZhangJianFeng2023}%
  \BibitemOpen
  \bibfield  {author} {\bibinfo {author} {\bibfnamefont {J.-F.}\ \bibnamefont
  {Zhang}}, \bibinfo {author} {\bibfnamefont {B.}~\bibnamefont {Zhan}},
  \bibinfo {author} {\bibfnamefont {M.}~\bibnamefont {Gao}}, \bibinfo {author}
  {\bibfnamefont {K.}~\bibnamefont {Liu}}, \bibinfo {author} {\bibfnamefont
  {X.}~\bibnamefont {Ren}}, \bibinfo {author} {\bibfnamefont {Z.-Y.}\
  \bibnamefont {Lu}}, \ and\ \bibinfo {author} {\bibfnamefont {T.}~\bibnamefont
  {Xiang}},\ }\href@noop {} {\bibfield  {journal} {\bibinfo  {journal} {Phys.
  Rev. B}\ }\textbf {\bibinfo {volume} {108}},\ \bibinfo {pages} {094519}
  (\bibinfo {year} {2023})}\BibitemShut {NoStop}%
\bibitem [{\citenamefont {Kortus}\ \emph {et~al.}(2001)\citenamefont {Kortus},
  \citenamefont {Mazin}, \citenamefont {Belashchenko}, \citenamefont
  {Antropov},\ and\ \citenamefont {Boyer}}]{mgb2_mazin}%
  \BibitemOpen
  \bibfield  {author} {\bibinfo {author} {\bibfnamefont {J.}~\bibnamefont
  {Kortus}}, \bibinfo {author} {\bibfnamefont {I.~I.}\ \bibnamefont {Mazin}},
  \bibinfo {author} {\bibfnamefont {K.~D.}\ \bibnamefont {Belashchenko}},
  \bibinfo {author} {\bibfnamefont {V.~P.}\ \bibnamefont {Antropov}}, \ and\
  \bibinfo {author} {\bibfnamefont {L.~L.}\ \bibnamefont {Boyer}},\ }\href
  {\doibase 10.1103/PhysRevLett.86.4656} {\bibfield  {journal} {\bibinfo
  {journal} {Phys. Rev. Lett.}\ }\textbf {\bibinfo {volume} {86}},\ \bibinfo
  {pages} {4656} (\bibinfo {year} {2001})}\BibitemShut {NoStop}%
\bibitem [{\citenamefont {Xi}(2008)}]{mgb2_xxx}%
  \BibitemOpen
  \bibfield  {author} {\bibinfo {author} {\bibfnamefont {X.~X.}\ \bibnamefont
  {Xi}},\ }\href {\doibase 10.1088/0034-4885/71/11/116501} {\bibfield
  {journal} {\bibinfo  {journal} {Rep. Prog. Phys.}\ }\textbf {\bibinfo
  {volume} {71}},\ \bibinfo {pages} {116501} (\bibinfo {year}
  {2008})}\BibitemShut {NoStop}%
\bibitem [{\citenamefont {Rosner}\ \emph {et~al.}(2002)\citenamefont {Rosner},
  \citenamefont {Kitaigorodsky},\ and\ \citenamefont
  {Pickett}}]{rosner2002prediction}%
  \BibitemOpen
  \bibfield  {author} {\bibinfo {author} {\bibfnamefont {H.}~\bibnamefont
  {Rosner}}, \bibinfo {author} {\bibfnamefont {A.}~\bibnamefont
  {Kitaigorodsky}}, \ and\ \bibinfo {author} {\bibfnamefont {W.}~\bibnamefont
  {Pickett}},\ }\href@noop {} {\bibfield  {journal} {\bibinfo  {journal} {Phys.
  Rev. Lett.}\ }\textbf {\bibinfo {volume} {88}},\ \bibinfo {pages} {127001}
  (\bibinfo {year} {2002})}\BibitemShut {NoStop}%
\bibitem [{\citenamefont {Dewhurst}\ \emph {et~al.}(2003)\citenamefont
  {Dewhurst}, \citenamefont {Sharma}, \citenamefont {Ambrosch-Draxl},\ and\
  \citenamefont {Johansson}}]{dewhurst2003first}%
  \BibitemOpen
  \bibfield  {author} {\bibinfo {author} {\bibfnamefont {J.}~\bibnamefont
  {Dewhurst}}, \bibinfo {author} {\bibfnamefont {S.}~\bibnamefont {Sharma}},
  \bibinfo {author} {\bibfnamefont {C.}~\bibnamefont {Ambrosch-Draxl}}, \ and\
  \bibinfo {author} {\bibfnamefont {B.}~\bibnamefont {Johansson}},\ }\href@noop
  {} {\bibfield  {journal} {\bibinfo  {journal} {Phys. Rev. B}\ }\textbf
  {\bibinfo {volume} {68}},\ \bibinfo {pages} {020504} (\bibinfo {year}
  {2003})}\BibitemShut {NoStop}%
\bibitem [{\citenamefont {Miao}\ \emph {et~al.}(2013)\citenamefont {Miao},
  \citenamefont {Yang}, \citenamefont {Jiang}, \citenamefont {Zhang},
  \citenamefont {Cai}, \citenamefont {Fan}, \citenamefont {Bai}, \citenamefont
  {Liu}, \citenamefont {Wu},\ and\ \citenamefont {Ma}}]{miao2013LiB1.1C0.9}%
  \BibitemOpen
  \bibfield  {author} {\bibinfo {author} {\bibfnamefont {R.}~\bibnamefont
  {Miao}}, \bibinfo {author} {\bibfnamefont {J.}~\bibnamefont {Yang}}, \bibinfo
  {author} {\bibfnamefont {M.}~\bibnamefont {Jiang}}, \bibinfo {author}
  {\bibfnamefont {Q.}~\bibnamefont {Zhang}}, \bibinfo {author} {\bibfnamefont
  {D.}~\bibnamefont {Cai}}, \bibinfo {author} {\bibfnamefont {C.}~\bibnamefont
  {Fan}}, \bibinfo {author} {\bibfnamefont {Z.}~\bibnamefont {Bai}}, \bibinfo
  {author} {\bibfnamefont {C.}~\bibnamefont {Liu}}, \bibinfo {author}
  {\bibfnamefont {F.}~\bibnamefont {Wu}}, \ and\ \bibinfo {author}
  {\bibfnamefont {S.}~\bibnamefont {Ma}},\ }\href@noop {} {\bibfield  {journal}
  {\bibinfo  {journal} {J. Appl. Phys.}\ }\textbf {\bibinfo {volume} {113}},\
  \bibinfo {pages} {133910} (\bibinfo {year} {2013})}\BibitemShut {NoStop}%
\bibitem [{\citenamefont {Bazhirov}\ \emph {et~al.}(2014)\citenamefont
  {Bazhirov}, \citenamefont {Sakai}, \citenamefont {Saito},\ and\ \citenamefont
  {Cohen}}]{bazhirov2014Li4B5C3}%
  \BibitemOpen
  \bibfield  {author} {\bibinfo {author} {\bibfnamefont {T.}~\bibnamefont
  {Bazhirov}}, \bibinfo {author} {\bibfnamefont {Y.}~\bibnamefont {Sakai}},
  \bibinfo {author} {\bibfnamefont {S.}~\bibnamefont {Saito}}, \ and\ \bibinfo
  {author} {\bibfnamefont {M.~L.}\ \bibnamefont {Cohen}},\ }\href@noop {}
  {\bibfield  {journal} {\bibinfo  {journal} {Phys. Rev. B}\ }\textbf {\bibinfo
  {volume} {89}},\ \bibinfo {pages} {045136} (\bibinfo {year}
  {2014})}\BibitemShut {NoStop}%
\bibitem [{\citenamefont {Li}\ \emph {et~al.}(2018)\citenamefont {Li},
  \citenamefont {Yan}, \citenamefont {Gao},\ and\ \citenamefont
  {Wang}}]{li2018LiB1+xC1-x}%
  \BibitemOpen
  \bibfield  {author} {\bibinfo {author} {\bibfnamefont {Q.-Z.}\ \bibnamefont
  {Li}}, \bibinfo {author} {\bibfnamefont {X.-W.}\ \bibnamefont {Yan}},
  \bibinfo {author} {\bibfnamefont {M.}~\bibnamefont {Gao}}, \ and\ \bibinfo
  {author} {\bibfnamefont {J.}~\bibnamefont {Wang}},\ }\href@noop {} {\bibfield
   {journal} {\bibinfo  {journal} {Euro. Phys. Lett.}\ }\textbf {\bibinfo
  {volume} {122}},\ \bibinfo {pages} {47001} (\bibinfo {year}
  {2018})}\BibitemShut {NoStop}%
\bibitem [{\citenamefont {Quan}\ and\ \citenamefont
  {Pickett}(2020)}]{quan_Pickett_2020li2Xbc3}%
  \BibitemOpen
  \bibfield  {author} {\bibinfo {author} {\bibfnamefont {Y.}~\bibnamefont
  {Quan}}\ and\ \bibinfo {author} {\bibfnamefont {W.~E.}\ \bibnamefont
  {Pickett}},\ }\href@noop {} {\bibfield  {journal} {\bibinfo  {journal} {Phys.
  Rev. B}\ }\textbf {\bibinfo {volume} {102}},\ \bibinfo {pages} {144504}
  (\bibinfo {year} {2020})}\BibitemShut {NoStop}%
\bibitem [{\citenamefont {Liu}\ \emph {et~al.}(2024)\citenamefont {Liu},
  \citenamefont {Wang}, \citenamefont {Fu}, \citenamefont {Lu},\ and\
  \citenamefont {Zhang}}]{liu2024three}%
  \BibitemOpen
  \bibfield  {author} {\bibinfo {author} {\bibfnamefont {H.-D.}\ \bibnamefont
  {Liu}}, \bibinfo {author} {\bibfnamefont {B.-T.}\ \bibnamefont {Wang}},
  \bibinfo {author} {\bibfnamefont {Z.-G.}\ \bibnamefont {Fu}}, \bibinfo
  {author} {\bibfnamefont {H.-Y.}\ \bibnamefont {Lu}}, \ and\ \bibinfo {author}
  {\bibfnamefont {P.}~\bibnamefont {Zhang}},\ }\href@noop {} {\bibfield
  {journal} {\bibinfo  {journal} {Phys. Rev. Res.}\ }\textbf {\bibinfo {volume}
  {6}},\ \bibinfo {pages} {033241} (\bibinfo {year} {2024})}\BibitemShut
  {NoStop}%
\bibitem [{\citenamefont {Yu}\ \emph {et~al.}(2024)\citenamefont {Yu},
  \citenamefont {Tang}, \citenamefont {Xu}, \citenamefont {Zhou}, \citenamefont
  {Ouyang}, \citenamefont {Deng}, \citenamefont {D'Agosta},\ and\ \citenamefont
  {Yang}}]{yu2024nontrivial}%
  \BibitemOpen
  \bibfield  {author} {\bibinfo {author} {\bibfnamefont {W.}~\bibnamefont
  {Yu}}, \bibinfo {author} {\bibfnamefont {J.-H.}\ \bibnamefont {Tang}},
  \bibinfo {author} {\bibfnamefont {H.-R.}\ \bibnamefont {Xu}}, \bibinfo
  {author} {\bibfnamefont {G.}~\bibnamefont {Zhou}}, \bibinfo {author}
  {\bibfnamefont {G.}~\bibnamefont {Ouyang}}, \bibinfo {author} {\bibfnamefont
  {H.-X.}\ \bibnamefont {Deng}}, \bibinfo {author} {\bibfnamefont
  {R.}~\bibnamefont {D'Agosta}}, \ and\ \bibinfo {author} {\bibfnamefont
  {K.}~\bibnamefont {Yang}},\ }\href@noop {} {\bibfield  {journal} {\bibinfo
  {journal} {New J. Phys.}\ } (\bibinfo {year} {2024})}\BibitemShut {NoStop}%
\bibitem [{\citenamefont {Pickett}(2006)}]{pickett2006design}%
  \BibitemOpen
  \bibfield  {author} {\bibinfo {author} {\bibfnamefont {W.~E.}\ \bibnamefont
  {Pickett}},\ }\href@noop {} {\bibfield  {journal} {\bibinfo  {journal} {J.
  Supercond. Nov. Mag.}\ }\textbf {\bibinfo {volume} {19}},\ \bibinfo {pages}
  {291} (\bibinfo {year} {2006})}\BibitemShut {NoStop}%
\bibitem [{\citenamefont {Pickett}(2017)}]{pickett2017revolution}%
  \BibitemOpen
  \bibfield  {author} {\bibinfo {author} {\bibfnamefont {W.~E.}\ \bibnamefont
  {Pickett}},\ }\href@noop {} {\bibfield  {journal} {\bibinfo  {journal} {arXiv
  preprint arXiv:1801.00165}\ } (\bibinfo {year} {2017})}\BibitemShut {NoStop}%
\bibitem [{\citenamefont {Fogg}\ \emph
  {et~al.}(2003{\natexlab{a}})\citenamefont {Fogg}, \citenamefont {Chalker},
  \citenamefont {Claridge}, \citenamefont {Darling},\ and\ \citenamefont
  {Rosseinsky}}]{fogg2003libc}%
  \BibitemOpen
  \bibfield  {author} {\bibinfo {author} {\bibfnamefont {A.}~\bibnamefont
  {Fogg}}, \bibinfo {author} {\bibfnamefont {P.}~\bibnamefont {Chalker}},
  \bibinfo {author} {\bibfnamefont {J.}~\bibnamefont {Claridge}}, \bibinfo
  {author} {\bibfnamefont {G.}~\bibnamefont {Darling}}, \ and\ \bibinfo
  {author} {\bibfnamefont {M.}~\bibnamefont {Rosseinsky}},\ }\href@noop {}
  {\bibfield  {journal} {\bibinfo  {journal} {Phys. Rev. B}\ }\textbf {\bibinfo
  {volume} {67}},\ \bibinfo {pages} {245106} (\bibinfo {year}
  {2003}{\natexlab{a}})}\BibitemShut {NoStop}%
\bibitem [{\citenamefont {Karimov}\ \emph {et~al.}(2004)\citenamefont
  {Karimov}, \citenamefont {Skorikov}, \citenamefont {Kurmaev}, \citenamefont
  {Finkelstein}, \citenamefont {Leitch}, \citenamefont {MacNaughton},
  \citenamefont {Moewes},\ and\ \citenamefont {Mori}}]{karimov2004libc}%
  \BibitemOpen
  \bibfield  {author} {\bibinfo {author} {\bibfnamefont {P.}~\bibnamefont
  {Karimov}}, \bibinfo {author} {\bibfnamefont {N.}~\bibnamefont {Skorikov}},
  \bibinfo {author} {\bibfnamefont {E.}~\bibnamefont {Kurmaev}}, \bibinfo
  {author} {\bibfnamefont {L.}~\bibnamefont {Finkelstein}}, \bibinfo {author}
  {\bibfnamefont {S.}~\bibnamefont {Leitch}}, \bibinfo {author} {\bibfnamefont
  {J.}~\bibnamefont {MacNaughton}}, \bibinfo {author} {\bibfnamefont
  {A.}~\bibnamefont {Moewes}}, \ and\ \bibinfo {author} {\bibfnamefont
  {T.}~\bibnamefont {Mori}},\ }\href@noop {} {\bibfield  {journal} {\bibinfo
  {journal} {J. Phys. Condens. Matter}\ }\textbf {\bibinfo {volume} {16}},\
  \bibinfo {pages} {5137} (\bibinfo {year} {2004})}\BibitemShut {NoStop}%
\bibitem [{\citenamefont {Bharathi}\ \emph {et~al.}(2002)\citenamefont
  {Bharathi}, \citenamefont {Balaselvi}, \citenamefont {Premila}, \citenamefont
  {Sairam}, \citenamefont {Reddy}, \citenamefont {Sundar},\ and\ \citenamefont
  {Hariharan}}]{bharathi2002synthesis}%
  \BibitemOpen
  \bibfield  {author} {\bibinfo {author} {\bibfnamefont {A.}~\bibnamefont
  {Bharathi}}, \bibinfo {author} {\bibfnamefont {S.~J.}\ \bibnamefont
  {Balaselvi}}, \bibinfo {author} {\bibfnamefont {M.}~\bibnamefont {Premila}},
  \bibinfo {author} {\bibfnamefont {T.}~\bibnamefont {Sairam}}, \bibinfo
  {author} {\bibfnamefont {G.}~\bibnamefont {Reddy}}, \bibinfo {author}
  {\bibfnamefont {C.}~\bibnamefont {Sundar}}, \ and\ \bibinfo {author}
  {\bibfnamefont {Y.}~\bibnamefont {Hariharan}},\ }\href@noop {} {\bibfield
  {journal} {\bibinfo  {journal} {Solid State Commun.}\ }\textbf {\bibinfo
  {volume} {124}},\ \bibinfo {pages} {423} (\bibinfo {year}
  {2002})}\BibitemShut {NoStop}%
\bibitem [{\citenamefont {Fogg}\ \emph
  {et~al.}(2003{\natexlab{b}})\citenamefont {Fogg}, \citenamefont {Claridge},
  \citenamefont {Darling},\ and\ \citenamefont
  {Rosseinsky}}]{fogg2003synthesis}%
  \BibitemOpen
  \bibfield  {author} {\bibinfo {author} {\bibfnamefont {A.}~\bibnamefont
  {Fogg}}, \bibinfo {author} {\bibfnamefont {J.}~\bibnamefont {Claridge}},
  \bibinfo {author} {\bibfnamefont {G.}~\bibnamefont {Darling}}, \ and\
  \bibinfo {author} {\bibfnamefont {M.}~\bibnamefont {Rosseinsky}},\
  }\href@noop {} {\bibfield  {journal} {\bibinfo  {journal} {Chem. Commun.}\ ,\
  \bibinfo {pages} {1348}} (\bibinfo {year} {2003}{\natexlab{b}})}\BibitemShut
  {NoStop}%
\bibitem [{\citenamefont {Souptel}\ \emph {et~al.}(2003)\citenamefont
  {Souptel}, \citenamefont {Hossain}, \citenamefont {Behr}, \citenamefont
  {L{\"o}ser},\ and\ \citenamefont {Geibel}}]{souptel2003synthesis}%
  \BibitemOpen
  \bibfield  {author} {\bibinfo {author} {\bibfnamefont {D.}~\bibnamefont
  {Souptel}}, \bibinfo {author} {\bibfnamefont {Z.}~\bibnamefont {Hossain}},
  \bibinfo {author} {\bibfnamefont {G.}~\bibnamefont {Behr}}, \bibinfo {author}
  {\bibfnamefont {W.}~\bibnamefont {L{\"o}ser}}, \ and\ \bibinfo {author}
  {\bibfnamefont {C.}~\bibnamefont {Geibel}},\ }\href@noop {} {\bibfield
  {journal} {\bibinfo  {journal} {Solid State Commun.}\ }\textbf {\bibinfo
  {volume} {125}},\ \bibinfo {pages} {17} (\bibinfo {year} {2003})}\BibitemShut
  {NoStop}%
\bibitem [{\citenamefont {Zhao}\ \emph {et~al.}(2003)\citenamefont {Zhao},
  \citenamefont {Klavins},\ and\ \citenamefont {Liu}}]{zhao2003synthesis}%
  \BibitemOpen
  \bibfield  {author} {\bibinfo {author} {\bibfnamefont {L.}~\bibnamefont
  {Zhao}}, \bibinfo {author} {\bibfnamefont {P.}~\bibnamefont {Klavins}}, \
  and\ \bibinfo {author} {\bibfnamefont {K.}~\bibnamefont {Liu}},\ }\href@noop
  {} {\bibfield  {journal} {\bibinfo  {journal} {J. Appl. Phys.}\ }\textbf
  {\bibinfo {volume} {93}},\ \bibinfo {pages} {8653} (\bibinfo {year}
  {2003})}\BibitemShut {NoStop}%
\bibitem [{\citenamefont {Nakamori}\ and\ \citenamefont
  {Orimo}(2004)}]{nakamori2004synthesis}%
  \BibitemOpen
  \bibfield  {author} {\bibinfo {author} {\bibfnamefont {Y.}~\bibnamefont
  {Nakamori}}\ and\ \bibinfo {author} {\bibfnamefont {S.-i.}\ \bibnamefont
  {Orimo}},\ }\href@noop {} {\bibfield  {journal} {\bibinfo  {journal} {J.
  Alloys Compd.}\ }\textbf {\bibinfo {volume} {370}},\ \bibinfo {pages} {L7}
  (\bibinfo {year} {2004})}\BibitemShut {NoStop}%
\bibitem [{\citenamefont {Fogg}\ \emph {et~al.}(2006)\citenamefont {Fogg},
  \citenamefont {Meldrum}, \citenamefont {Darling}, \citenamefont {Claridge},\
  and\ \citenamefont {Rosseinsky}}]{fogg2006chemical}%
  \BibitemOpen
  \bibfield  {author} {\bibinfo {author} {\bibfnamefont {A.~M.}\ \bibnamefont
  {Fogg}}, \bibinfo {author} {\bibfnamefont {J.}~\bibnamefont {Meldrum}},
  \bibinfo {author} {\bibfnamefont {G.~R.}\ \bibnamefont {Darling}}, \bibinfo
  {author} {\bibfnamefont {J.~B.}\ \bibnamefont {Claridge}}, \ and\ \bibinfo
  {author} {\bibfnamefont {M.~J.}\ \bibnamefont {Rosseinsky}},\ }\href@noop {}
  {\bibfield  {journal} {\bibinfo  {journal} {J. Am. Chem. Soc.}\ }\textbf
  {\bibinfo {volume} {128}},\ \bibinfo {pages} {10043} (\bibinfo {year}
  {2006})}\BibitemShut {NoStop}%
\bibitem [{\citenamefont {Burdett}\ \emph {et~al.}(1985)\citenamefont
  {Burdett}, \citenamefont {Lee},\ and\ \citenamefont
  {McLarnan}}]{burdett1985coloring}%
  \BibitemOpen
  \bibfield  {author} {\bibinfo {author} {\bibfnamefont {J.~K.}\ \bibnamefont
  {Burdett}}, \bibinfo {author} {\bibfnamefont {S.}~\bibnamefont {Lee}}, \ and\
  \bibinfo {author} {\bibfnamefont {T.~J.}\ \bibnamefont {McLarnan}},\
  }\href@noop {} {\bibfield  {journal} {\bibinfo  {journal} {J. Am. Chem.
  Soc.}\ }\textbf {\bibinfo {volume} {107}},\ \bibinfo {pages} {3083} (\bibinfo
  {year} {1985})}\BibitemShut {NoStop}%
\bibitem [{\citenamefont {Miller}(1998)}]{miller1998coloring}%
  \BibitemOpen
  \bibfield  {author} {\bibinfo {author} {\bibfnamefont {G.~J.}\ \bibnamefont
  {Miller}},\ }\href@noop {} {\bibfield  {journal} {\bibinfo  {journal} {Euro.
  J. Inorg. Chem.}\ ,\ \bibinfo {pages} {523}} (\bibinfo {year}
  {1998})}\BibitemShut {NoStop}%
\bibitem [{\citenamefont {Sanchez}\ \emph {et~al.}(1984)\citenamefont
  {Sanchez}, \citenamefont {Ducastelle},\ and\ \citenamefont
  {Gratias}}]{sanchez1984generalized}%
  \BibitemOpen
  \bibfield  {author} {\bibinfo {author} {\bibfnamefont {J.~M.}\ \bibnamefont
  {Sanchez}}, \bibinfo {author} {\bibfnamefont {F.}~\bibnamefont {Ducastelle}},
  \ and\ \bibinfo {author} {\bibfnamefont {D.}~\bibnamefont {Gratias}},\
  }\href@noop {} {\bibfield  {journal} {\bibinfo  {journal} {Phys. A: Stat.
  Mech. Appl.}\ }\textbf {\bibinfo {volume} {128}},\ \bibinfo {pages} {334}
  (\bibinfo {year} {1984})}\BibitemShut {NoStop}%
\bibitem [{\citenamefont {De~Fontaine}(1994)}]{Fontaine1994}%
  \BibitemOpen
  \bibfield  {author} {\bibinfo {author} {\bibfnamefont {D.}~\bibnamefont
  {De~Fontaine}},\ }\href@noop {} {\bibfield  {journal} {\bibinfo  {journal}
  {Solid State Phys.}\ }\textbf {\bibinfo {volume} {47}},\ \bibinfo {pages}
  {33} (\bibinfo {year} {1994})}\BibitemShut {NoStop}%
\bibitem [{\citenamefont {Zunger}(1994)}]{Zunger1994}%
  \BibitemOpen
  \bibfield  {author} {\bibinfo {author} {\bibfnamefont {A.}~\bibnamefont
  {Zunger}},\ }\enquote {\bibinfo {title} {First-principles statistical
  mechanics of semiconductor alloys and intermetallic compounds},}\ in\
  \href@noop {} {\emph {\bibinfo {booktitle} {Statics and Dynamics of Alloy
  Phase Transformations}}}\ (\bibinfo  {publisher} {Springer},\ \bibinfo {year}
  {1994})\ pp.\ \bibinfo {pages} {361--419}\BibitemShut {NoStop}%
\bibitem [{\citenamefont {Giustino}\ \emph {et~al.}(2007)\citenamefont
  {Giustino}, \citenamefont {Cohen},\ and\ \citenamefont
  {Louie}}]{Giustino2007}%
  \BibitemOpen
  \bibfield  {author} {\bibinfo {author} {\bibfnamefont {F.}~\bibnamefont
  {Giustino}}, \bibinfo {author} {\bibfnamefont {M.~L.}\ \bibnamefont {Cohen}},
  \ and\ \bibinfo {author} {\bibfnamefont {S.~G.}\ \bibnamefont {Louie}},\
  }\href@noop {} {\bibfield  {journal} {\bibinfo  {journal} {Phys. Rev. B}\
  }\textbf {\bibinfo {volume} {76}},\ \bibinfo {pages} {165108} (\bibinfo
  {year} {2007})}\BibitemShut {NoStop}%
\bibitem [{\citenamefont {Ponc{\'e}}\ \emph {et~al.}(2016)\citenamefont
  {Ponc{\'e}}, \citenamefont {Margine}, \citenamefont {Verdi},\ and\
  \citenamefont {Giustino}}]{ponce2016epw}%
  \BibitemOpen
  \bibfield  {author} {\bibinfo {author} {\bibfnamefont {S.}~\bibnamefont
  {Ponc{\'e}}}, \bibinfo {author} {\bibfnamefont {E.~R.}\ \bibnamefont
  {Margine}}, \bibinfo {author} {\bibfnamefont {C.}~\bibnamefont {Verdi}}, \
  and\ \bibinfo {author} {\bibfnamefont {F.}~\bibnamefont {Giustino}},\
  }\href@noop {} {\bibfield  {journal} {\bibinfo  {journal} {Comput. Phys.
  Commun.}\ }\textbf {\bibinfo {volume} {209}},\ \bibinfo {pages} {116}
  (\bibinfo {year} {2016})}\BibitemShut {NoStop}%
\bibitem [{\citenamefont {Lee}\ \emph {et~al.}(2023)\citenamefont {Lee},
  \citenamefont {Ponce}, \citenamefont {Bushick}, \citenamefont {Hajinazar},
  \citenamefont {Lafuente-Bartolome}, \citenamefont {Leveillee}, \citenamefont
  {Lian}, \citenamefont {Lihm}, \citenamefont {Macheda}, \citenamefont {Mori},
  \citenamefont {Paudyal}, \citenamefont {Sio}, \citenamefont {Tiwari},
  \citenamefont {Zacharias}, \citenamefont {Zhang}, \citenamefont {Bonini},
  \citenamefont {Kioupakis}, \citenamefont {Margine},\ and\ \citenamefont
  {Giustino}}]{Lee2023epw}%
  \BibitemOpen
  \bibfield  {author} {\bibinfo {author} {\bibfnamefont {H.}~\bibnamefont
  {Lee}}, \bibinfo {author} {\bibfnamefont {S.}~\bibnamefont {Ponce}}, \bibinfo
  {author} {\bibfnamefont {K.}~\bibnamefont {Bushick}}, \bibinfo {author}
  {\bibfnamefont {S.}~\bibnamefont {Hajinazar}}, \bibinfo {author}
  {\bibfnamefont {J.}~\bibnamefont {Lafuente-Bartolome}}, \bibinfo {author}
  {\bibfnamefont {J.}~\bibnamefont {Leveillee}}, \bibinfo {author}
  {\bibfnamefont {C.}~\bibnamefont {Lian}}, \bibinfo {author} {\bibfnamefont
  {J.-M.}\ \bibnamefont {Lihm}}, \bibinfo {author} {\bibfnamefont
  {F.}~\bibnamefont {Macheda}}, \bibinfo {author} {\bibfnamefont
  {H.}~\bibnamefont {Mori}}, \bibinfo {author} {\bibfnamefont {H.}~\bibnamefont
  {Paudyal}}, \bibinfo {author} {\bibfnamefont {W.~H.}\ \bibnamefont {Sio}},
  \bibinfo {author} {\bibfnamefont {S.}~\bibnamefont {Tiwari}}, \bibinfo
  {author} {\bibfnamefont {M.}~\bibnamefont {Zacharias}}, \bibinfo {author}
  {\bibfnamefont {X.}~\bibnamefont {Zhang}}, \bibinfo {author} {\bibfnamefont
  {N.}~\bibnamefont {Bonini}}, \bibinfo {author} {\bibfnamefont
  {E.}~\bibnamefont {Kioupakis}}, \bibinfo {author} {\bibfnamefont {E.~R.}\
  \bibnamefont {Margine}}, \ and\ \bibinfo {author} {\bibfnamefont
  {F.}~\bibnamefont {Giustino}},\ }\href {\doibase 10.1038/s41524-023-01107-3}
  {\bibfield  {journal} {\bibinfo  {journal} {Npj Comput. Mater.}\ }\textbf
  {\bibinfo {volume} {9}} (\bibinfo {year} {2023}),\
  10.1038/s41524-023-01107-3}\BibitemShut {NoStop}%
\bibitem [{see()}]{see_appendix}%
  \BibitemOpen
  \href@noop {} {}\bibinfo {note} {Please find more details in our
  appendix.}\BibitemShut {Stop}%
\bibitem [{\citenamefont {Allen}(1972)}]{allen1972neutron}%
  \BibitemOpen
  \bibfield  {author} {\bibinfo {author} {\bibfnamefont {P.~B.}\ \bibnamefont
  {Allen}},\ }\href@noop {} {\bibfield  {journal} {\bibinfo  {journal} {Phys.
  Rev. B}\ }\textbf {\bibinfo {volume} {6}},\ \bibinfo {pages} {2577} (\bibinfo
  {year} {1972})}\BibitemShut {NoStop}%
\bibitem [{\citenamefont {Allen}\ and\ \citenamefont
  {Dynes}(1975)}]{allen1975transition}%
  \BibitemOpen
  \bibfield  {author} {\bibinfo {author} {\bibfnamefont {P.~B.}\ \bibnamefont
  {Allen}}\ and\ \bibinfo {author} {\bibfnamefont {R.}~\bibnamefont {Dynes}},\
  }\href@noop {} {\bibfield  {journal} {\bibinfo  {journal} {Phys. Rev. B}\
  }\textbf {\bibinfo {volume} {12}},\ \bibinfo {pages} {905} (\bibinfo {year}
  {1975})}\BibitemShut {NoStop}%
\bibitem [{\citenamefont {Khan}\ and\ \citenamefont
  {Allen}(1984)}]{khan1984deformation}%
  \BibitemOpen
  \bibfield  {author} {\bibinfo {author} {\bibfnamefont {F.}~\bibnamefont
  {Khan}}\ and\ \bibinfo {author} {\bibfnamefont {P.}~\bibnamefont {Allen}},\
  }\href@noop {} {\bibfield  {journal} {\bibinfo  {journal} {Phys. Rev. B}\
  }\textbf {\bibinfo {volume} {29}},\ \bibinfo {pages} {3341} (\bibinfo {year}
  {1984})}\BibitemShut {NoStop}%
\bibitem [{\citenamefont {Lee}\ and\ \citenamefont
  {Pickett}(2005)}]{lee2005crystal}%
  \BibitemOpen
  \bibfield  {author} {\bibinfo {author} {\bibfnamefont {K.-W.}\ \bibnamefont
  {Lee}}\ and\ \bibinfo {author} {\bibfnamefont {W.}~\bibnamefont {Pickett}},\
  }\href@noop {} {\bibfield  {journal} {\bibinfo  {journal} {Phys. Rev. B}\
  }\textbf {\bibinfo {volume} {72}},\ \bibinfo {pages} {174505} (\bibinfo
  {year} {2005})}\BibitemShut {NoStop}%
\bibitem [{\citenamefont {Tan}\ \emph {et~al.}(2021)\citenamefont {Tan},
  \citenamefont {Liu}, \citenamefont {Wang},\ and\ \citenamefont
  {Yan}}]{tan2021dp}%
  \BibitemOpen
  \bibfield  {author} {\bibinfo {author} {\bibfnamefont {H.}~\bibnamefont
  {Tan}}, \bibinfo {author} {\bibfnamefont {Y.}~\bibnamefont {Liu}}, \bibinfo
  {author} {\bibfnamefont {Z.}~\bibnamefont {Wang}}, \ and\ \bibinfo {author}
  {\bibfnamefont {B.}~\bibnamefont {Yan}},\ }\href@noop {} {\bibfield
  {journal} {\bibinfo  {journal} {Phys. Rev. Lett.}\ }\textbf {\bibinfo
  {volume} {127}},\ \bibinfo {pages} {046401} (\bibinfo {year}
  {2021})}\BibitemShut {NoStop}%
\bibitem [{\citenamefont {Hoffmann}(1991)}]{hoffmann1991COOP}%
  \BibitemOpen
  \bibfield  {author} {\bibinfo {author} {\bibfnamefont {R.}~\bibnamefont
  {Hoffmann}},\ }\href@noop {} {\emph {\bibinfo {title} {Solids and surfaces: a
  chemist's view of bonding in extended structures}}}\ (\bibinfo  {publisher}
  {John Wiley \& Sons},\ \bibinfo {year} {1991})\BibitemShut {NoStop}%
\bibitem [{\citenamefont {Dronskowski}\ and\ \citenamefont
  {Bl{\"o}chl}(1993)}]{dronskowski1993COHP}%
  \BibitemOpen
  \bibfield  {author} {\bibinfo {author} {\bibfnamefont {R.}~\bibnamefont
  {Dronskowski}}\ and\ \bibinfo {author} {\bibfnamefont {P.~E.}\ \bibnamefont
  {Bl{\"o}chl}},\ }\href@noop {} {\bibfield  {journal} {\bibinfo  {journal} {J.
  Phys. Chem.}\ }\textbf {\bibinfo {volume} {97}},\ \bibinfo {pages} {8617}
  (\bibinfo {year} {1993})}\BibitemShut {NoStop}%
\bibitem [{\citenamefont {Sano}\ \emph {et~al.}(2016)\citenamefont {Sano},
  \citenamefont {Koretsune}, \citenamefont {Tadano}, \citenamefont {Akashi},\
  and\ \citenamefont {Arita}}]{sano2016VHS}%
  \BibitemOpen
  \bibfield  {author} {\bibinfo {author} {\bibfnamefont {W.}~\bibnamefont
  {Sano}}, \bibinfo {author} {\bibfnamefont {T.}~\bibnamefont {Koretsune}},
  \bibinfo {author} {\bibfnamefont {T.}~\bibnamefont {Tadano}}, \bibinfo
  {author} {\bibfnamefont {R.}~\bibnamefont {Akashi}}, \ and\ \bibinfo {author}
  {\bibfnamefont {R.}~\bibnamefont {Arita}},\ }\href@noop {} {\bibfield
  {journal} {\bibinfo  {journal} {Phys. Rev. B}\ }\textbf {\bibinfo {volume}
  {93}},\ \bibinfo {pages} {094525} (\bibinfo {year} {2016})}\BibitemShut
  {NoStop}%
\bibitem [{\citenamefont {Gai}\ \emph {et~al.}(2022)\citenamefont {Gai},
  \citenamefont {Guo}, \citenamefont {Yang}, \citenamefont {Gao}, \citenamefont
  {Gao},\ and\ \citenamefont {Lu}}]{gai2022B3C3}%
  \BibitemOpen
  \bibfield  {author} {\bibinfo {author} {\bibfnamefont {T.-T.}\ \bibnamefont
  {Gai}}, \bibinfo {author} {\bibfnamefont {P.-J.}\ \bibnamefont {Guo}},
  \bibinfo {author} {\bibfnamefont {H.-C.}\ \bibnamefont {Yang}}, \bibinfo
  {author} {\bibfnamefont {Y.}~\bibnamefont {Gao}}, \bibinfo {author}
  {\bibfnamefont {M.}~\bibnamefont {Gao}}, \ and\ \bibinfo {author}
  {\bibfnamefont {Z.-Y.}\ \bibnamefont {Lu}},\ }\href@noop {} {\bibfield
  {journal} {\bibinfo  {journal} {Phys. Rev. B}\ }\textbf {\bibinfo {volume}
  {105}},\ \bibinfo {pages} {224514} (\bibinfo {year} {2022})}\BibitemShut
  {NoStop}%
\bibitem [{\citenamefont {Gerber}\ \emph {et~al.}(2017)\citenamefont {Gerber},
  \citenamefont {Yang}, \citenamefont {Zhu}, \citenamefont {Soifer},
  \citenamefont {Sobota}, \citenamefont {Rebec}, \citenamefont {Lee},
  \citenamefont {Jia}, \citenamefont {Moritz}, \citenamefont {Jia} \emph
  {et~al.}}]{ZXShen2017elph_dp}%
  \BibitemOpen
  \bibfield  {author} {\bibinfo {author} {\bibfnamefont {S.}~\bibnamefont
  {Gerber}}, \bibinfo {author} {\bibfnamefont {S.-L.}\ \bibnamefont {Yang}},
  \bibinfo {author} {\bibfnamefont {D.}~\bibnamefont {Zhu}}, \bibinfo {author}
  {\bibfnamefont {H.}~\bibnamefont {Soifer}}, \bibinfo {author} {\bibfnamefont
  {J.}~\bibnamefont {Sobota}}, \bibinfo {author} {\bibfnamefont
  {S.}~\bibnamefont {Rebec}}, \bibinfo {author} {\bibfnamefont
  {J.}~\bibnamefont {Lee}}, \bibinfo {author} {\bibfnamefont {T.}~\bibnamefont
  {Jia}}, \bibinfo {author} {\bibfnamefont {B.}~\bibnamefont {Moritz}},
  \bibinfo {author} {\bibfnamefont {C.}~\bibnamefont {Jia}},  \emph {et~al.},\
  }\href@noop {} {\bibfield  {journal} {\bibinfo  {journal} {Science}\ }\textbf
  {\bibinfo {volume} {357}},\ \bibinfo {pages} {71} (\bibinfo {year}
  {2017})}\BibitemShut {NoStop}%
\bibitem [{\citenamefont {Huang}\ \emph {et~al.}(2023)\citenamefont {Huang},
  \citenamefont {Querales-Flores}, \citenamefont {Teitelbaum}, \citenamefont
  {Cao}, \citenamefont {Henighan}, \citenamefont {Liu}, \citenamefont {Jiang},
  \citenamefont {De~la Pe{\~n}a}, \citenamefont {Krapivin}, \citenamefont
  {Haber} \emph {et~al.}}]{huang2023ultrafast}%
  \BibitemOpen
  \bibfield  {author} {\bibinfo {author} {\bibfnamefont {Y.}~\bibnamefont
  {Huang}}, \bibinfo {author} {\bibfnamefont {J.~D.}\ \bibnamefont
  {Querales-Flores}}, \bibinfo {author} {\bibfnamefont {S.~W.}\ \bibnamefont
  {Teitelbaum}}, \bibinfo {author} {\bibfnamefont {J.}~\bibnamefont {Cao}},
  \bibinfo {author} {\bibfnamefont {T.}~\bibnamefont {Henighan}}, \bibinfo
  {author} {\bibfnamefont {H.}~\bibnamefont {Liu}}, \bibinfo {author}
  {\bibfnamefont {M.}~\bibnamefont {Jiang}}, \bibinfo {author} {\bibfnamefont
  {G.}~\bibnamefont {De~la Pe{\~n}a}}, \bibinfo {author} {\bibfnamefont
  {V.}~\bibnamefont {Krapivin}}, \bibinfo {author} {\bibfnamefont
  {J.}~\bibnamefont {Haber}},  \emph {et~al.},\ }\href@noop {} {\bibfield
  {journal} {\bibinfo  {journal} {Phys. Rev. X}\ }\textbf {\bibinfo {volume}
  {13}},\ \bibinfo {pages} {041050} (\bibinfo {year} {2023})}\BibitemShut
  {NoStop}%
\bibitem [{\citenamefont {Sobota}\ \emph {et~al.}(2021)\citenamefont {Sobota},
  \citenamefont {He},\ and\ \citenamefont {Shen}}]{sobota2021angle}%
  \BibitemOpen
  \bibfield  {author} {\bibinfo {author} {\bibfnamefont {J.~A.}\ \bibnamefont
  {Sobota}}, \bibinfo {author} {\bibfnamefont {Y.}~\bibnamefont {He}}, \ and\
  \bibinfo {author} {\bibfnamefont {Z.-X.}\ \bibnamefont {Shen}},\ }\href@noop
  {} {\bibfield  {journal} {\bibinfo  {journal} {Rev. Mod. Phys.}\ }\textbf
  {\bibinfo {volume} {93}},\ \bibinfo {pages} {025006} (\bibinfo {year}
  {2021})}\BibitemShut {NoStop}%
\bibitem [{\citenamefont {Boschini}\ \emph {et~al.}(2024)\citenamefont
  {Boschini}, \citenamefont {Zonno},\ and\ \citenamefont
  {Damascelli}}]{boschini2024time}%
  \BibitemOpen
  \bibfield  {author} {\bibinfo {author} {\bibfnamefont {F.}~\bibnamefont
  {Boschini}}, \bibinfo {author} {\bibfnamefont {M.}~\bibnamefont {Zonno}}, \
  and\ \bibinfo {author} {\bibfnamefont {A.}~\bibnamefont {Damascelli}},\
  }\href@noop {} {\bibfield  {journal} {\bibinfo  {journal} {Rev. Mod. Phys.}\
  }\textbf {\bibinfo {volume} {96}},\ \bibinfo {pages} {015003} (\bibinfo
  {year} {2024})}\BibitemShut {NoStop}%
\bibitem [{\citenamefont {Bhaumik}\ \emph {et~al.}(2017)\citenamefont
  {Bhaumik}, \citenamefont {Sachan},\ and\ \citenamefont
  {Narayan}}]{bhaumik2017high}%
  \BibitemOpen
  \bibfield  {author} {\bibinfo {author} {\bibfnamefont {A.}~\bibnamefont
  {Bhaumik}}, \bibinfo {author} {\bibfnamefont {R.}~\bibnamefont {Sachan}}, \
  and\ \bibinfo {author} {\bibfnamefont {J.}~\bibnamefont {Narayan}},\
  }\href@noop {} {\bibfield  {journal} {\bibinfo  {journal} {ACS nano}\
  }\textbf {\bibinfo {volume} {11}},\ \bibinfo {pages} {5351} (\bibinfo {year}
  {2017})}\BibitemShut {NoStop}%
\bibitem [{\citenamefont {Sun}\ \emph {et~al.}(2023)\citenamefont {Sun},
  \citenamefont {Huo}, \citenamefont {Hu}, \citenamefont {Li}, \citenamefont
  {Liu}, \citenamefont {Han}, \citenamefont {Tang}, \citenamefont {Mao},
  \citenamefont {Yang}, \citenamefont {Wang}, \citenamefont {Cheng},
  \citenamefont {Yao}, \citenamefont {Zhang},\ and\ \citenamefont
  {Wang}}]{sun2023}%
  \BibitemOpen
  \bibfield  {author} {\bibinfo {author} {\bibfnamefont {H.}~\bibnamefont
  {Sun}}, \bibinfo {author} {\bibfnamefont {M.}~\bibnamefont {Huo}}, \bibinfo
  {author} {\bibfnamefont {X.}~\bibnamefont {Hu}}, \bibinfo {author}
  {\bibfnamefont {J.}~\bibnamefont {Li}}, \bibinfo {author} {\bibfnamefont
  {Z.}~\bibnamefont {Liu}}, \bibinfo {author} {\bibfnamefont {Y.}~\bibnamefont
  {Han}}, \bibinfo {author} {\bibfnamefont {L.}~\bibnamefont {Tang}}, \bibinfo
  {author} {\bibfnamefont {Z.}~\bibnamefont {Mao}}, \bibinfo {author}
  {\bibfnamefont {P.}~\bibnamefont {Yang}}, \bibinfo {author} {\bibfnamefont
  {B.}~\bibnamefont {Wang}}, \bibinfo {author} {\bibfnamefont {J.}~\bibnamefont
  {Cheng}}, \bibinfo {author} {\bibfnamefont {D.-X.}\ \bibnamefont {Yao}},
  \bibinfo {author} {\bibfnamefont {G.-M.}\ \bibnamefont {Zhang}}, \ and\
  \bibinfo {author} {\bibfnamefont {M.}~\bibnamefont {Wang}},\ }\href {\doibase
  10.1038/s41586-023-06408-7} {\bibfield  {journal} {\bibinfo  {journal}
  {Nature}\ }\textbf {\bibinfo {volume} {621}},\ \bibinfo {pages} {493}
  (\bibinfo {year} {2023})}\BibitemShut {NoStop}%
\bibitem [{\citenamefont {Ding}\ \emph {et~al.}(2023)\citenamefont {Ding},
  \citenamefont {Tam}, \citenamefont {Sui}, \citenamefont {Zhao}, \citenamefont
  {Xu}, \citenamefont {Choi}, \citenamefont {Leng}, \citenamefont {Zhang},
  \citenamefont {Wu}, \citenamefont {Xiao} \emph {et~al.}}]{ding2023critical}%
  \BibitemOpen
  \bibfield  {author} {\bibinfo {author} {\bibfnamefont {X.}~\bibnamefont
  {Ding}}, \bibinfo {author} {\bibfnamefont {C.~C.}\ \bibnamefont {Tam}},
  \bibinfo {author} {\bibfnamefont {X.}~\bibnamefont {Sui}}, \bibinfo {author}
  {\bibfnamefont {Y.}~\bibnamefont {Zhao}}, \bibinfo {author} {\bibfnamefont
  {M.}~\bibnamefont {Xu}}, \bibinfo {author} {\bibfnamefont {J.}~\bibnamefont
  {Choi}}, \bibinfo {author} {\bibfnamefont {H.}~\bibnamefont {Leng}}, \bibinfo
  {author} {\bibfnamefont {J.}~\bibnamefont {Zhang}}, \bibinfo {author}
  {\bibfnamefont {M.}~\bibnamefont {Wu}}, \bibinfo {author} {\bibfnamefont
  {H.}~\bibnamefont {Xiao}},  \emph {et~al.},\ }\href@noop {} {\bibfield
  {journal} {\bibinfo  {journal} {Nature}\ }\textbf {\bibinfo {volume} {615}},\
  \bibinfo {pages} {50} (\bibinfo {year} {2023})}\BibitemShut {NoStop}%
\bibitem [{\citenamefont {Van De~Walle}(2009)}]{vdWalle2009}%
  \BibitemOpen
  \bibfield  {author} {\bibinfo {author} {\bibfnamefont {A.}~\bibnamefont {Van
  De~Walle}},\ }\href@noop {} {\bibfield  {journal} {\bibinfo  {journal}
  {Calphad}\ }\textbf {\bibinfo {volume} {33}},\ \bibinfo {pages} {266}
  (\bibinfo {year} {2009})}\BibitemShut {NoStop}%
\bibitem [{\citenamefont {Van~de Walle}\ and\ \citenamefont
  {Ceder}(2002)}]{vdWalle2002}%
  \BibitemOpen
  \bibfield  {author} {\bibinfo {author} {\bibfnamefont {A.}~\bibnamefont
  {Van~de Walle}}\ and\ \bibinfo {author} {\bibfnamefont {G.}~\bibnamefont
  {Ceder}},\ }\href@noop {} {\bibfield  {journal} {\bibinfo  {journal} {J.
  Phase Equilib.}\ }\textbf {\bibinfo {volume} {23}},\ \bibinfo {pages} {348}
  (\bibinfo {year} {2002})}\BibitemShut {NoStop}%
\bibitem [{\citenamefont {Baroni}\ \emph {et~al.}(2001)\citenamefont {Baroni},
  \citenamefont {De~Gironcoli}, \citenamefont {Dal~Corso},\ and\ \citenamefont
  {Giannozzi}}]{baroni2001phonons}%
  \BibitemOpen
  \bibfield  {author} {\bibinfo {author} {\bibfnamefont {S.}~\bibnamefont
  {Baroni}}, \bibinfo {author} {\bibfnamefont {S.}~\bibnamefont
  {De~Gironcoli}}, \bibinfo {author} {\bibfnamefont {A.}~\bibnamefont
  {Dal~Corso}}, \ and\ \bibinfo {author} {\bibfnamefont {P.}~\bibnamefont
  {Giannozzi}},\ }\href@noop {} {\bibfield  {journal} {\bibinfo  {journal}
  {Rev. Mod. Phys.}\ }\textbf {\bibinfo {volume} {73}},\ \bibinfo {pages} {515}
  (\bibinfo {year} {2001})}\BibitemShut {NoStop}%
\bibitem [{\citenamefont {Kresse}\ and\ \citenamefont
  {Furthm{\"u}ller}(1996)}]{kresse1996}%
  \BibitemOpen
  \bibfield  {author} {\bibinfo {author} {\bibfnamefont {G.}~\bibnamefont
  {Kresse}}\ and\ \bibinfo {author} {\bibfnamefont {J.}~\bibnamefont
  {Furthm{\"u}ller}},\ }\href@noop {} {\bibfield  {journal} {\bibinfo
  {journal} {Phys. Rev. B}\ }\textbf {\bibinfo {volume} {54}},\ \bibinfo
  {pages} {11169} (\bibinfo {year} {1996})}\BibitemShut {NoStop}%
\bibitem [{\citenamefont {Kresse}\ and\ \citenamefont
  {Joubert}(1999)}]{Joubert1999}%
  \BibitemOpen
  \bibfield  {author} {\bibinfo {author} {\bibfnamefont {G.}~\bibnamefont
  {Kresse}}\ and\ \bibinfo {author} {\bibfnamefont {D.}~\bibnamefont
  {Joubert}},\ }\href@noop {} {\bibfield  {journal} {\bibinfo  {journal} {Phys.
  Rev. B}\ }\textbf {\bibinfo {volume} {59}},\ \bibinfo {pages} {1758}
  (\bibinfo {year} {1999})}\BibitemShut {NoStop}%
\bibitem [{\citenamefont {Perdew}\ \emph {et~al.}(2008)\citenamefont {Perdew},
  \citenamefont {Ruzsinszky}, \citenamefont {Csonka}, \citenamefont {Vydrov},
  \citenamefont {Scuseria}, \citenamefont {Constantin}, \citenamefont {Zhou},\
  and\ \citenamefont {Burke}}]{perdew2008}%
  \BibitemOpen
  \bibfield  {author} {\bibinfo {author} {\bibfnamefont {J.~P.}\ \bibnamefont
  {Perdew}}, \bibinfo {author} {\bibfnamefont {A.}~\bibnamefont {Ruzsinszky}},
  \bibinfo {author} {\bibfnamefont {G.~I.}\ \bibnamefont {Csonka}}, \bibinfo
  {author} {\bibfnamefont {O.~A.}\ \bibnamefont {Vydrov}}, \bibinfo {author}
  {\bibfnamefont {G.~E.}\ \bibnamefont {Scuseria}}, \bibinfo {author}
  {\bibfnamefont {L.~A.}\ \bibnamefont {Constantin}}, \bibinfo {author}
  {\bibfnamefont {X.}~\bibnamefont {Zhou}}, \ and\ \bibinfo {author}
  {\bibfnamefont {K.}~\bibnamefont {Burke}},\ }\href@noop {} {\bibfield
  {journal} {\bibinfo  {journal} {Phys. Rev. Lett.}\ }\textbf {\bibinfo
  {volume} {100}},\ \bibinfo {pages} {136406} (\bibinfo {year}
  {2008})}\BibitemShut {NoStop}%
\bibitem [{\citenamefont {Van De~Walle}\ \emph {et~al.}(2002)\citenamefont {Van
  De~Walle}, \citenamefont {Asta},\ and\ \citenamefont {Ceder}}]{vdWalle2002c}%
  \BibitemOpen
  \bibfield  {author} {\bibinfo {author} {\bibfnamefont {A.}~\bibnamefont {Van
  De~Walle}}, \bibinfo {author} {\bibfnamefont {M.}~\bibnamefont {Asta}}, \
  and\ \bibinfo {author} {\bibfnamefont {G.}~\bibnamefont {Ceder}},\
  }\href@noop {} {\bibfield  {journal} {\bibinfo  {journal} {Calphad}\ }\textbf
  {\bibinfo {volume} {26}},\ \bibinfo {pages} {539} (\bibinfo {year}
  {2002})}\BibitemShut {NoStop}%
\bibitem [{\citenamefont {Bl{\"o}chl}(1994)}]{blochl1994projector}%
  \BibitemOpen
  \bibfield  {author} {\bibinfo {author} {\bibfnamefont {P.~E.}\ \bibnamefont
  {Bl{\"o}chl}},\ }\href@noop {} {\bibfield  {journal} {\bibinfo  {journal}
  {Phys. Rev. B}\ }\textbf {\bibinfo {volume} {50}},\ \bibinfo {pages} {17953}
  (\bibinfo {year} {1994})}\BibitemShut {NoStop}%
\bibitem [{\citenamefont {Mostofi}\ \emph {et~al.}(2008)\citenamefont
  {Mostofi}, \citenamefont {Yates}, \citenamefont {Lee}, \citenamefont {Souza},
  \citenamefont {Vanderbilt},\ and\ \citenamefont
  {Marzari}}]{mostofi2008wannier90}%
  \BibitemOpen
  \bibfield  {author} {\bibinfo {author} {\bibfnamefont {A.~A.}\ \bibnamefont
  {Mostofi}}, \bibinfo {author} {\bibfnamefont {J.~R.}\ \bibnamefont {Yates}},
  \bibinfo {author} {\bibfnamefont {Y.-S.}\ \bibnamefont {Lee}}, \bibinfo
  {author} {\bibfnamefont {I.}~\bibnamefont {Souza}}, \bibinfo {author}
  {\bibfnamefont {D.}~\bibnamefont {Vanderbilt}}, \ and\ \bibinfo {author}
  {\bibfnamefont {N.}~\bibnamefont {Marzari}},\ }\href@noop {} {\bibfield
  {journal} {\bibinfo  {journal} {Comput. Phys. Commun.}\ }\textbf {\bibinfo
  {volume} {178}},\ \bibinfo {pages} {685} (\bibinfo {year}
  {2008})}\BibitemShut {NoStop}%
\bibitem [{\citenamefont {Marzari}\ \emph {et~al.}(2012)\citenamefont
  {Marzari}, \citenamefont {Mostofi}, \citenamefont {Yates}, \citenamefont
  {Souza},\ and\ \citenamefont {Vanderbilt}}]{Marzari2012}%
  \BibitemOpen
  \bibfield  {author} {\bibinfo {author} {\bibfnamefont {N.}~\bibnamefont
  {Marzari}}, \bibinfo {author} {\bibfnamefont {A.~A.}\ \bibnamefont
  {Mostofi}}, \bibinfo {author} {\bibfnamefont {J.~R.}\ \bibnamefont {Yates}},
  \bibinfo {author} {\bibfnamefont {I.}~\bibnamefont {Souza}}, \ and\ \bibinfo
  {author} {\bibfnamefont {D.}~\bibnamefont {Vanderbilt}},\ }\href@noop {}
  {\bibfield  {journal} {\bibinfo  {journal} {Rev. Mod. Phys.}\ ,\ \bibinfo
  {pages} {1419}} (\bibinfo {year} {2012})}\BibitemShut {NoStop}%
\bibitem [{\citenamefont {Giannozzi}\ \emph {et~al.}(2009)\citenamefont
  {Giannozzi}, \citenamefont {Baroni}, \citenamefont {Bonini}, \citenamefont
  {Calandra}, \citenamefont {Car}, \citenamefont {Cavazzoni}, \citenamefont
  {Ceresoli}, \citenamefont {Chiarotti}, \citenamefont {Cococcioni},
  \citenamefont {Dabo} \emph {et~al.}}]{giannozzi2009quantum}%
  \BibitemOpen
  \bibfield  {author} {\bibinfo {author} {\bibfnamefont {P.}~\bibnamefont
  {Giannozzi}}, \bibinfo {author} {\bibfnamefont {S.}~\bibnamefont {Baroni}},
  \bibinfo {author} {\bibfnamefont {N.}~\bibnamefont {Bonini}}, \bibinfo
  {author} {\bibfnamefont {M.}~\bibnamefont {Calandra}}, \bibinfo {author}
  {\bibfnamefont {R.}~\bibnamefont {Car}}, \bibinfo {author} {\bibfnamefont
  {C.}~\bibnamefont {Cavazzoni}}, \bibinfo {author} {\bibfnamefont
  {D.}~\bibnamefont {Ceresoli}}, \bibinfo {author} {\bibfnamefont {G.~L.}\
  \bibnamefont {Chiarotti}}, \bibinfo {author} {\bibfnamefont {M.}~\bibnamefont
  {Cococcioni}}, \bibinfo {author} {\bibfnamefont {I.}~\bibnamefont {Dabo}},
  \emph {et~al.},\ }\href@noop {} {\bibfield  {journal} {\bibinfo  {journal}
  {J. Phys. Condens. Matter}\ }\textbf {\bibinfo {volume} {21}},\ \bibinfo
  {pages} {395502} (\bibinfo {year} {2009})}\BibitemShut {NoStop}%
\bibitem [{\citenamefont {Perdew}\ \emph {et~al.}(1996)\citenamefont {Perdew},
  \citenamefont {Burke},\ and\ \citenamefont {Ernzerhof}}]{perdew1996}%
  \BibitemOpen
  \bibfield  {author} {\bibinfo {author} {\bibfnamefont {J.~P.}\ \bibnamefont
  {Perdew}}, \bibinfo {author} {\bibfnamefont {K.}~\bibnamefont {Burke}}, \
  and\ \bibinfo {author} {\bibfnamefont {M.}~\bibnamefont {Ernzerhof}},\
  }\href@noop {} {\bibfield  {journal} {\bibinfo  {journal} {Phys. Rev. Lett.}\
  }\textbf {\bibinfo {volume} {77}},\ \bibinfo {pages} {3865} (\bibinfo {year}
  {1996})}\BibitemShut {NoStop}%
\bibitem [{\citenamefont {Troullier}\ and\ \citenamefont
  {Martins}(1991)}]{Normconserving}%
  \BibitemOpen
  \bibfield  {author} {\bibinfo {author} {\bibfnamefont {N.}~\bibnamefont
  {Troullier}}\ and\ \bibinfo {author} {\bibfnamefont {J.~L.}\ \bibnamefont
  {Martins}},\ }\href@noop {} {\bibfield  {journal} {\bibinfo  {journal} {Phys.
  Rev. B}\ }\textbf {\bibinfo {volume} {43}},\ \bibinfo {pages} {1993}
  (\bibinfo {year} {1991})}\BibitemShut {NoStop}%
\bibitem [{\citenamefont {Hamann}(2013)}]{hamann2013ONCV}%
  \BibitemOpen
  \bibfield  {author} {\bibinfo {author} {\bibfnamefont {D.}~\bibnamefont
  {Hamann}},\ }\href@noop {} {\bibfield  {journal} {\bibinfo  {journal} {Phys.
  Rev. B}\ }\textbf {\bibinfo {volume} {88}},\ \bibinfo {pages} {085117}
  (\bibinfo {year} {2013})}\BibitemShut {NoStop}%
\bibitem [{\citenamefont {Schlipf}\ and\ \citenamefont
  {Gygi}(2015)}]{schlipf2015sg15}%
  \BibitemOpen
  \bibfield  {author} {\bibinfo {author} {\bibfnamefont {M.}~\bibnamefont
  {Schlipf}}\ and\ \bibinfo {author} {\bibfnamefont {F.}~\bibnamefont {Gygi}},\
  }\href@noop {} {\bibfield  {journal} {\bibinfo  {journal} {Comput. Phys.
  Commun.}\ }\textbf {\bibinfo {volume} {196}},\ \bibinfo {pages} {36}
  (\bibinfo {year} {2015})}\BibitemShut {NoStop}%
\bibitem [{\citenamefont {Marzari}\ \emph {et~al.}(1999)\citenamefont
  {Marzari}, \citenamefont {Vanderbilt}, \citenamefont {De~Vita},\ and\
  \citenamefont {Payne}}]{coldsmearing}%
  \BibitemOpen
  \bibfield  {author} {\bibinfo {author} {\bibfnamefont {N.}~\bibnamefont
  {Marzari}}, \bibinfo {author} {\bibfnamefont {D.}~\bibnamefont {Vanderbilt}},
  \bibinfo {author} {\bibfnamefont {A.}~\bibnamefont {De~Vita}}, \ and\
  \bibinfo {author} {\bibfnamefont {M.}~\bibnamefont {Payne}},\ }\href@noop {}
  {\bibfield  {journal} {\bibinfo  {journal} {Phys. Rev. Lett.}\ }\textbf
  {\bibinfo {volume} {82}},\ \bibinfo {pages} {3296} (\bibinfo {year}
  {1999})}\BibitemShut {NoStop}%
\bibitem [{\citenamefont {Setyawan}\ and\ \citenamefont
  {Curtarolo}(2010)}]{aflowkpath}%
  \BibitemOpen
  \bibfield  {author} {\bibinfo {author} {\bibfnamefont {W.}~\bibnamefont
  {Setyawan}}\ and\ \bibinfo {author} {\bibfnamefont {S.}~\bibnamefont
  {Curtarolo}},\ }\href@noop {} {\bibfield  {journal} {\bibinfo  {journal}
  {Comput. Mater. Sci.}\ }\textbf {\bibinfo {volume} {49}},\ \bibinfo {pages}
  {299} (\bibinfo {year} {2010})}\BibitemShut {NoStop}%
\end{thebibliography}%

\clearpage
\appendix

\section{Theoretical methods and computational details}
\subsection{the cluster expansion method}
We use the cluster expansion method \cite{sanchez1984generalized} to calculate the configuration dependence of the energy of alloys. In the cluster expansion method applied here to the binary system \ce{Li_nB_{n+1}C_{n-1}}, we select the \ce{AlB2} lattice as the parent lattice and defined the configuration $\sigma$ by specifying the occupations of each of the $n$ honeycomb lattice sites by a B atom or a C atom. For each configuration, the pseudo-spin variables, $\hat{S}_i(i=1,2,\ldots,n)$, are assigned to each of the $n$ sites with $\hat{S}_i=-1$ or $+1$ depending on the site i occupied by a B atom or a C atom, respectively. Then any configuration $\sigma$ can be represented as a vector of pseudo-spin variables and its energy can be mapped onto a generalized Ising Hamiltonian \cite{Fontaine1994,Zunger1994}
\matheq{eq:F-CE-2}{
  E(\bm{\sigma}) = N \sum_\alpha m_\alpha J_\alpha \overline{\Pi}_{\alpha}(\bm{\sigma}),
}
where the coefficients $J_0$, $J_i$, $J_{ij}$,$\cdots$ are termed as effective cluster interactions (ECIs), $\alpha$ denotes the clusters that are symmetrically distinct, $m_\alpha$ is the multiplicity of the cluster $\alpha$ per lattice site, $J_\alpha$ is the Ising-type interaction energy for the cluster $\alpha$. $\overline{\Pi}_{\alpha}(\sigma)$ is the lattice-averaged correlation function defined as a product of the variables $\hat{S}_i$ for all sites of cluster $\alpha$ and the overbar denotes an average over all symmetry equivalent clusters
\matheq{eq:corr-func}{
  \overline{\Pi}_{\alpha}(\bm{\sigma}) =  \frac{1}{N m_\alpha} \sum_{(i_1<i_2<\cdots<i_g)\in \alpha} \hat{S}_{i_1} \hat{S}_{i_2} \cdots \hat{S}_{i_g},
}
where the summation is over all the symmetrically equivalent clusters under the space group operations  of the lattice. The completeness and orthonormality of the set $\lbrace\Pi_{\alpha}(\bm{\sigma})\rbrace$ of pseudo-spin products in the space of all configurations assure that expansions in Eq. \ref{eq:F-CE-2} are exact if all $2^n$ figures are used to describe all the $2^n$ configurational energies. In practice, the summation considered in the cluster expansion is readily truncated to a finite number of clusters including pairs, triplets and few quadruplets. The ECI parameters are determined by fitting the target properties of tens of representative configurations in their relaxed geometric structures calculated by first-principles approaches \cite{Zunger1994, vdWalle2009} . The transferability of ECIs can be measured by the leave-one-out cross validation (LOOCV) score \cite{vdWalle2002},
\begin{equation}
    \label{eq:cv}
    S_{CV}= \sqrt{\frac{1}{N_c}\sum^{N_c}_s (E_{s}-\hat{E}_{s})^2},
\end{equation}
where $N_c$ is the number of configurations used for fitting, $E_{s}$ denotes the physical quantity calculated from the first principles approach, and $\hat{E}_{s}$ is the predicted value from cluster expansion with the ECI parameters fitted by the remaining $N_c - 1$ configurations.

\subsection{the calculation of phonon-mediated superconductivity}
The superconducting transition temperature $T_c$ is calculated with the McMillan-Allen-Dynes fomula \cite{allen1972neutron,allen1975transition}:
\begin{equation}
    T_c=\frac{\omega_\mathrm{log}}{1.2}{\rm exp}[\frac{-1.04(1+\lambda)}{\lambda(1-0.62\mu^*)-\mu^*}].
\end{equation}
Here $\lambda$ is the total EPC strength, $\mu^*$ is a parameter as the effect screened Coulomb pseudopotential and $\omega_\mathrm{log}$ is the logarithmic average of the Eliashberg spectral function. $\lambda$ and $\omega_\mathrm{log}$ are defined as:
\begin{align}
\lambda&=\sum_{\textbf{q}\nu}\lambda_{\textbf{q}\nu}=2\int\frac{\alpha^2F(\omega)}{\omega}\mathrm{d}\omega, \\
\omega_\mathrm{log}&={\rm exp}[\frac{2}{\lambda}\int\frac{\mathrm{d}\omega}{\omega}\alpha^2F(\omega)ln(\omega)],\\
\alpha^2F(\omega)&=\frac{1}{2}\sum_{\textbf{q}\nu}\delta(\omega-\omega_{\textbf{q}\nu})\omega_{\textbf{q}\nu}\lambda_{\textbf{q}\nu}.
\end{align}
Here $\omega_{\textbf{q}\nu}$ and $ \lambda_{\textbf{q}\nu}$ are the frequency and EPC strength of the $\nu$-th phonon mode at the wave vector $\textbf{q}$. The $\lambda_{\textbf{q}\nu}$ is defined as:
\begin{eqnarray}
    \lambda_{\vb{q}\nu}=\dfrac{\gamma_{\vb{q}\nu}}{\pi\hbar N(0)\omega^2_{\vb{q}\nu}}.
\end{eqnarray}
Here $\gamma_{\vb{q}\nu}$ is the phonon linewidth:
\begin{equation}
\gamma_{\vb{q}\nu}=2\pi\omega_{\vb{q}\nu}\sum_{i,j}\int\dfrac{\mathrm{d}\vb{k}}{\Omega_{BZ}}|g_{\vb{q}\nu}(\vb{k},i,j)|^2\delta(\epsilon_{\vb{k},i}-\epsilon_F)\delta(\epsilon_{\vb{k}+\vb{q},j}-\epsilon_F).
\end{equation}
Here $i/j$ is the band index. The electron-phonon interaction matrix element $g_{\vb{q}\nu}(\vb{k},i,j)$, describing the probability amplitude for scattering of an electron with a transfer of crystal momentum $\vb{q}$, is determined by:
\begin{eqnarray}\label{elph-matrix-def}
    g_{\vb{q}\nu}(\vb{k},i,j)&=&\sqrt{\dfrac{\hbar}{2\omega_{\vb{q}\nu}}}<\psi_{\vb{k},i}|\Delta V^{\vb{q}\nu}_{KS}|\psi_{\vb{k}+\vb{q},j}>; \nonumber \\
    \Delta V^{\vb{q}\nu}_{KS}&=&\sum_{\vb{R},s}\pdv{V_{KS}}{\vb{u}_{s\vb{R}}}\vb{u}^{\vb{q}\nu}_s\dfrac{e^{i\vb{qR}}}{\sqrt{N}}
\end{eqnarray}
Here $\vb{u}^{\vb{q}\nu}_s=\dfrac{\vb{\nu}^{\vb{q}\nu}_s}{\sqrt{M_s}}$ and $\pdv{V_{KS}}{\vb{u}_{s\vb{R}}}$ can be calculated by density functional perturbation theory (DFPT) \cite{baroni2001phonons}.

\subsection{computational details}

In order to build cluster expansion of the energy, we firstly employ VASP \cite{kresse1996,Joubert1999} to perform DFT calculations for different configurations with certain \ce{AlB2}-type supercells with PBEsol functional \cite{perdew2008}. The cluster expansions are constructed using the alloy theoretic automated toolkit (ATAT) \cite{vdWalle2002c}. To fit ECI with DFT results, we consider different randomly generated configurations in the $2\times2\times2$ and the $2\times3\times2$ \ce{AlB2}-type supercells with different proportion of boron to carbon. For each configuration, both the lattice constants and internal coordinates are fully relaxed while the energy convergence criterion is $10^{-6}$ eV and the force convergence criterion is 0.01 eV/\AA. The projector augmented wave \cite{blochl1994projector} (PAW) approach is employed and the cut-off energy of the plane wave basis is 600 eV. The Brillouin zones in supercells are sampled by $\mathbf{k}$-grids as dense as a $\Gamma$-centered $6\times6\times6$ $\mathbf{k}$-grid in an \ce{AlB2}-type unit cell. LOOCVs of both \ce{Li2B3C} and \ce{Li3B4C2} are smaller than 0.01 eV.

We employ Wannier90 \cite{mostofi2008wannier90,Marzari2012} interfacing with Quantum ESPRESSO \cite{giannozzi2009quantum} to calculate maximally localized WFs for analyzing electronic structures. We also calculate the lattice constants and total energy with Quantum ESPRESSO. We adopt PBE exchange-correlation functional \cite{perdew1996} and the SG15 norm-conserving pseudopotentials \cite{Normconserving,hamann2013ONCV,schlipf2015sg15}. The crystal structures are fully relaxed while the energy convergence criterion is $10^{-12}$ Ry and the force convergence criterion is $10^{-6}$ Ry/Bohr. The kinetic energy cutoff is set to 80 Ry. The Marzari-Vanderbilt-DeVita-Payne smearing method \cite{coldsmearing} is employed with the spreading of 0.015 Ry. The $\mathbf{k}$-grid is $16\times16\times12$/$16\times16\times6$/$16\times8\times6$/$16\times16\times6$/$12\times12\times12$/$9\times9\times12$ for \ce{MgB2}/\ce{LiBC}/\ce{Li2B3C}(TW)/\ce{Li2B3C}(GLX)/\ce{Li3B4C2} (TW)/\ce{Li3B4C2}(GLX), respectively. The energy windows and frozen windows for the disentanglement procedure in Wannierization are listed in our appendix. The Wannier centers are set on B/C atoms for $p_z$-like WFs and on B-B/B-C bond center for $s$-like B-B/B-C bonding WFs.

We employ EPW package \cite{ponce2016epw,Lee2023epw} to calculate the electron-phonon coupling properties of \ce{Li2B3C} (TW) and \ce{Li3B4C2} (GLX). We take the $16\times8\times6$/$9\times9\times12$ $\mathbf{k}$-mesh and $8\times4\times3$/$3\times3\times4$ $\textbf{q}$-mesh as coarse grids and then interpolate them to the dense $64\times32\times24$/$36\times36\times48$ $\mathbf{k}$-mesh and $16\times8\times6$/$9\times9\times12$ $\textbf{q}$-mesh in \ce{Li2B3C} (TW)/\ce{Li3B4C2} (GLX), respectively. The Gaussian smearing method with the width of 0.05 Ry is used for the Fermi surface broadening.

\clearpage

\section{Cluster expansions of L\lowercase{i}$_2$B$_{3}$C and L\lowercase{i}$_4$B$_{4}$C$_2$}
Our cluster expansion results can faithfully predict the energies of different configurations of \ce{Li_nB_{n+1}C_{n-1}}.
\begin{figure}[htb]
	 \centerline{\includegraphics[width=0.5\textwidth]{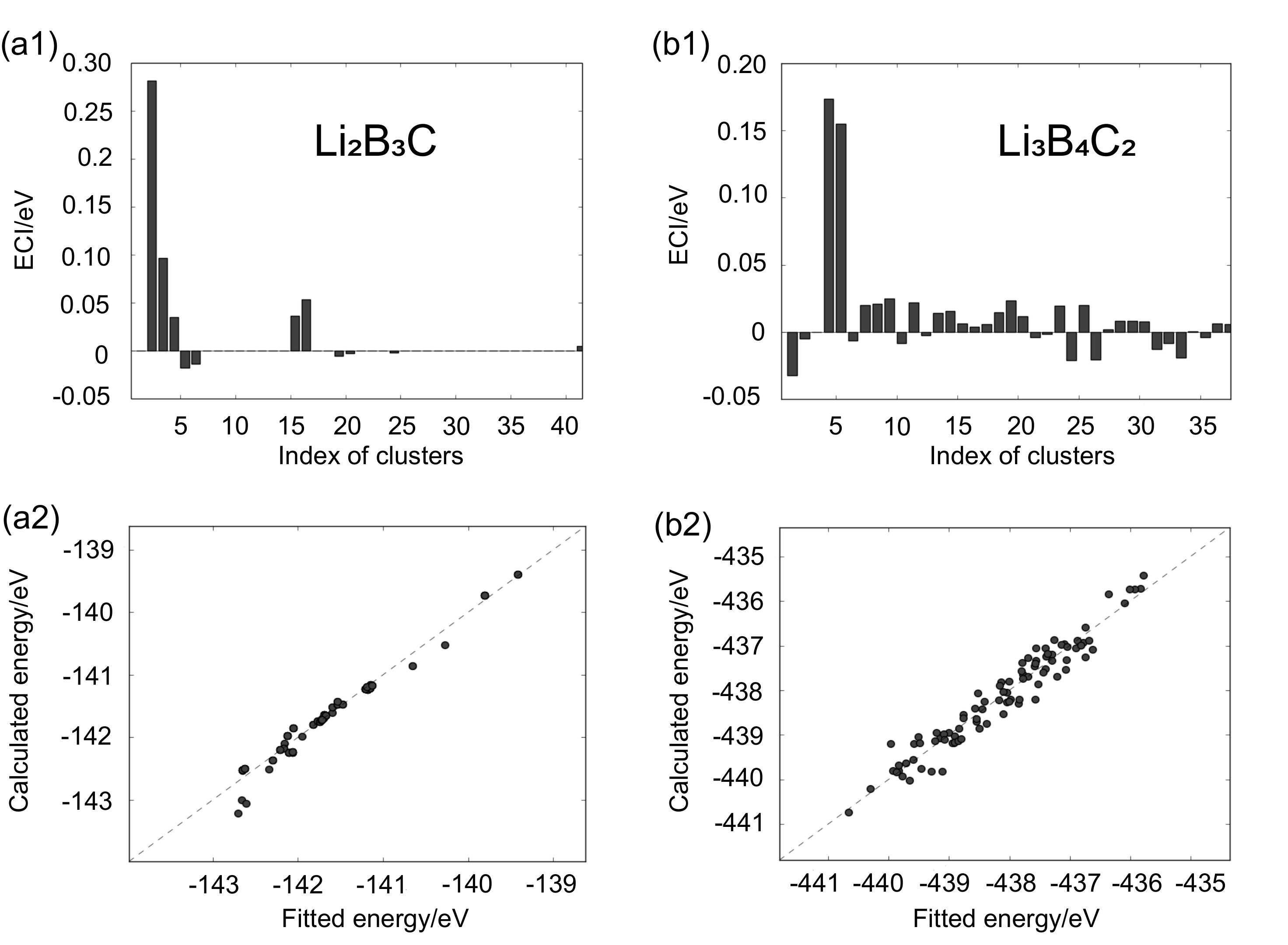}}
 \caption{(a1) ECIs for cluster expansion of \ce{Li2B3C}. (a2) The DFT-calculated and cluster-expansion-fitted energies of \ce{Li2B3C}. (b1) ECIs for cluster expansion of \ce{Li3B4C2}. (b2) The DFT-calculated and cluster-expansion-fitted energies of \ce{Li3B4C2}.
		\label{ce} }
\end{figure}

\clearpage

\section{The high-symmetry \textbf{k}-points in band calculations}

\begin{figure}[htb]
	 \centerline{\includegraphics[width=0.5\textwidth]{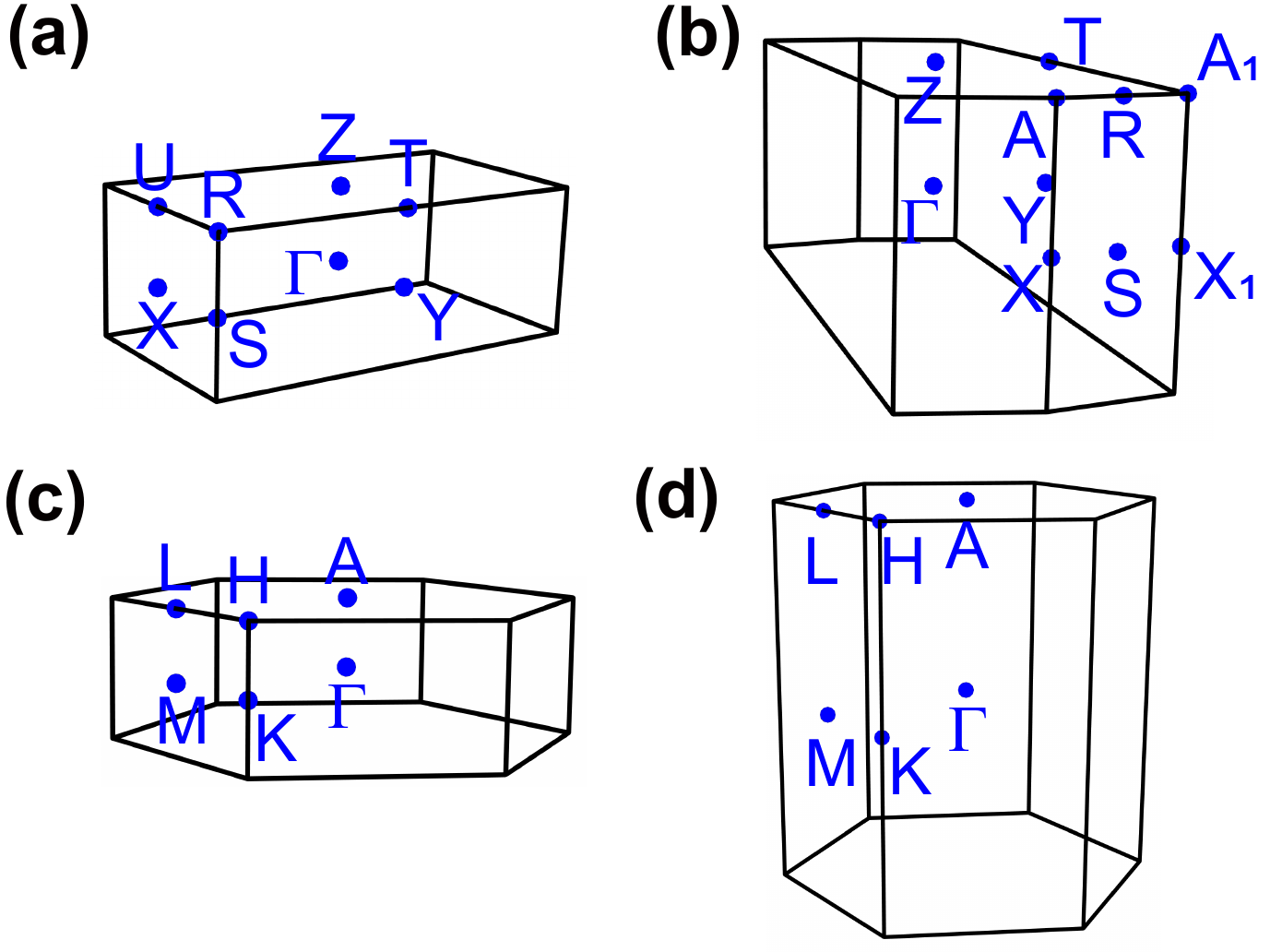}}
 \caption{The first Brillouin zones and the high-symmetry \textbf{k}-points of (a) \ce{Li2B3C} (TW), (b) \ce{Li3B4C2} (TW), (c) \ce{Li2B3C} (GLX), and (d) \ce{Li3B4C2} (GLX).
		\label{kpoints} }
\end{figure}

The first Brillouin zones and the high-symmetry \textbf{k}-points in main text are shown in Fig. \ref{kpoints}. The choices of the high-symmetry \textbf{k}-points are same as ALFOW's \cite{aflowkpath}.

\clearpage

\section{details of the electronic structures of B$_{n+1}$C$_{n-1}$ layers}
\begin{figure*}[htb]
	\centerline{\includegraphics[width=0.8\textwidth]{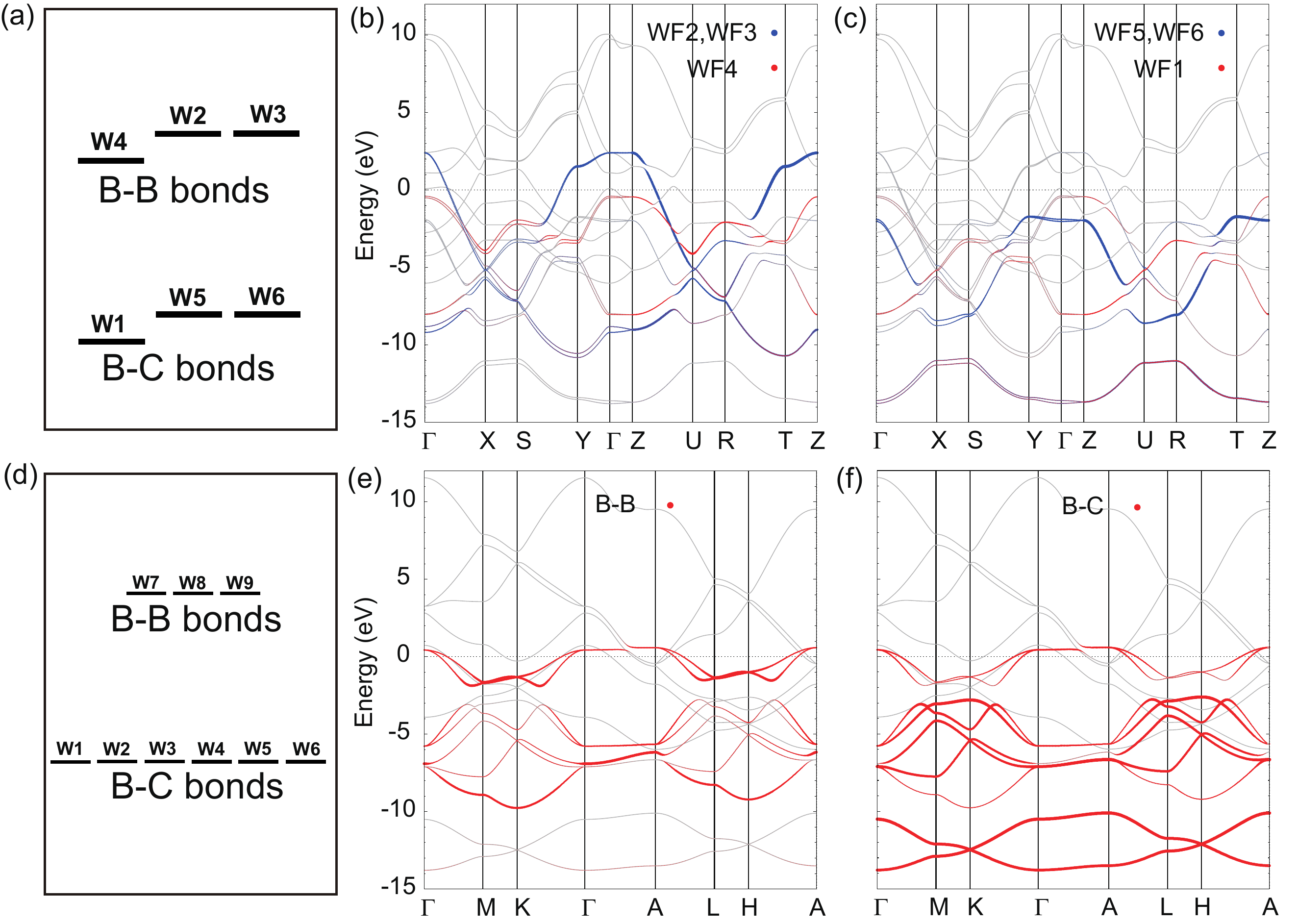}} \caption{(a) The schematic illustration of the on-site energies of B-B/B-C $\sigma$-bonding WFs. (b-c) The band structure of \ce{Li2B3C} (TW) with projection on B-B/B-C $\sigma$-bonding WFs. (d) The schematic illustration of the on-site energies of B-B/B-C $\sigma$-bonding WFs. (e-f) The band structure of \ce{Li3B4C2} (GLX) with projection on B-B/B-C $\sigma$-bonding WFs.
		\label{band_hybridize} }
\end{figure*}

Here we show the details of electronic structures of the \ce{B3C} layer in \ce{Li2B3C} (TW) and  the \ce{B4C2} layer in \ce{Li3B4C2} (GLX).  In \ce{Li3B4C2} (GLX), the B-B bonding bands strongly hybridize with the B-C bonding bands across the energy range of -10 eV to 1 eV, making it difficult to distinguish. To demonstrate this clearly, we perform fat band analyses for the B-B bonding bands and B-C bonding bands separately, as shown in Fig. \ref{band_hybridize}(e-f).

However, in \ce{Li2B3C} (TW), the B-B $\sigma$-bonding bands from the B-B zigzag chain cross the Fermi level and weakly hybridize with the B-C bonding bands from the B-C zigzag chain, due to the significant difference in energy levels of the B-B and B-C $\sigma$-bonding WFs and the absence of nearest-neighbour hopping between B-B and B-C zigzag chains. As a result, we can clearly distinguish the bands from the B-B/B-C zigzag chains. These bands are quite dispersive along their corresponding extending direction, as shown in Fig. \ref{band_hybridize}(b-c). The shape of Fermi surfaces also changes owing to the band dispersion, different from the cylindrical Fermi surface in \ce{MgB2} \cite{mgb2_mazin}, as shown in our main text. The electronic physics is similar in \ce{Li3B4C2} (TW): the B-B $\sigma$-bonding bands cross the Fermi level and weakly couple with the B-C $\sigma$-bonding bands.

In contrast, the B-B $\sigma$ bonds can not form infinitely linked B-B zigzag chains with other B-B $\sigma$ bonds in \ce{Li3B4C2} (GLX): the triangle formed by three B-B $\sigma$ bonds is surrounded by other B-C $\sigma$ bonds, as shown in Fig. \ref{fig1}(d). Consequently, the B-B $\sigma$-bonding bands are forced to hybridize with B-C $\sigma$-bonding bands, as shown in Fig. \ref{band_hybridize}(e-f).

\clearpage

\section{Prototypes: MgB$_2$ and LiBC}

\begin{figure}[t] \centerline{\includegraphics[width=0.5\textwidth]{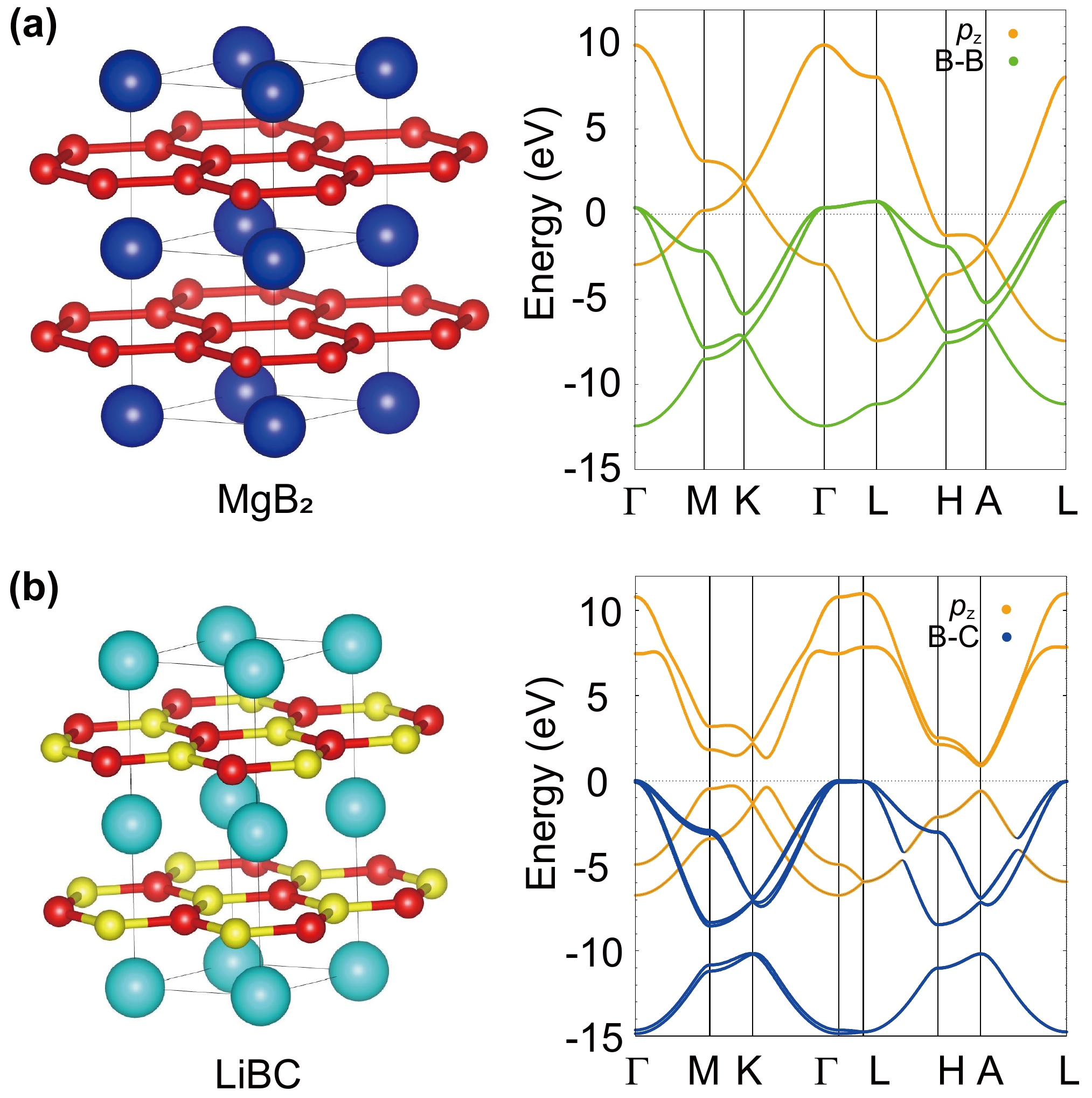}}
 \caption{Crystal structures and band structures of (a) \ce{MgB2} and (b) LiBC. The sizes of dots are proportional to the spectral weights. 
 		\label{prototypes} }
\end{figure}
To have a holistic view for the electronic structures of \ce{LinB_{n+1}C_{n-1}}, we firstly revisit the prototypes \ce{MgB2} and LiBC. \ce{MgB2} is a textbook example of realizing high-temperature phonon-mediated superconductivity by metallizing the $\sigma$-bonding electrons. The crystal structure of \ce{MgB2} is depicted in Fig. \ref{prototypes}(a). In \ce{MgB2}, the crystal field effect induced by the \ce{Mg^{2+}} ion results in metallic $sp^2$ $\sigma$-bonding electrons. Consequently, holes from the $\pi$ bands are transferred into these $\sigma$-bonding bands, positioning the Fermi level near the upper boundary of the $\sigma$-bonding bands, as illustrated in Fig. \ref{prototypes}(a). Using aforementioned method to analyse its doping level, the B-B $\sigma$-bonding bands in \ce{MgB2} are nearly fully occupied without considering self-doping from $\pi$-bands.

The lattice structure of LiBC has been established experimentally, as depicted in Fig. \ref{prototypes}(b): the B and C atoms in LiBC form an ordered pattern, alternating at the vertices of a hexagonal lattice within each BC layer. The disparity in the on-site energies of the B and C $p_z$ orbitals gives rise to an energy gap, rendering LiBC an insulator with properties similar to boron nitride (BN), as shown in Fig. \ref{prototypes}(b). The energy gap in the $\pi$ band precludes the partial occupation of the bonding $\sigma$ bands. In next session, We will conclude the features of the electronic structures of \ce{MgB2}, LiBC and \ce{LinB_{n+1}C_{n-1}} to have a holistic view.

\clearpage

\section{Simplified physical pictures of electronic structures}

\begin{figure}[htb]
	\centerline{\includegraphics[width=0.5\textwidth]{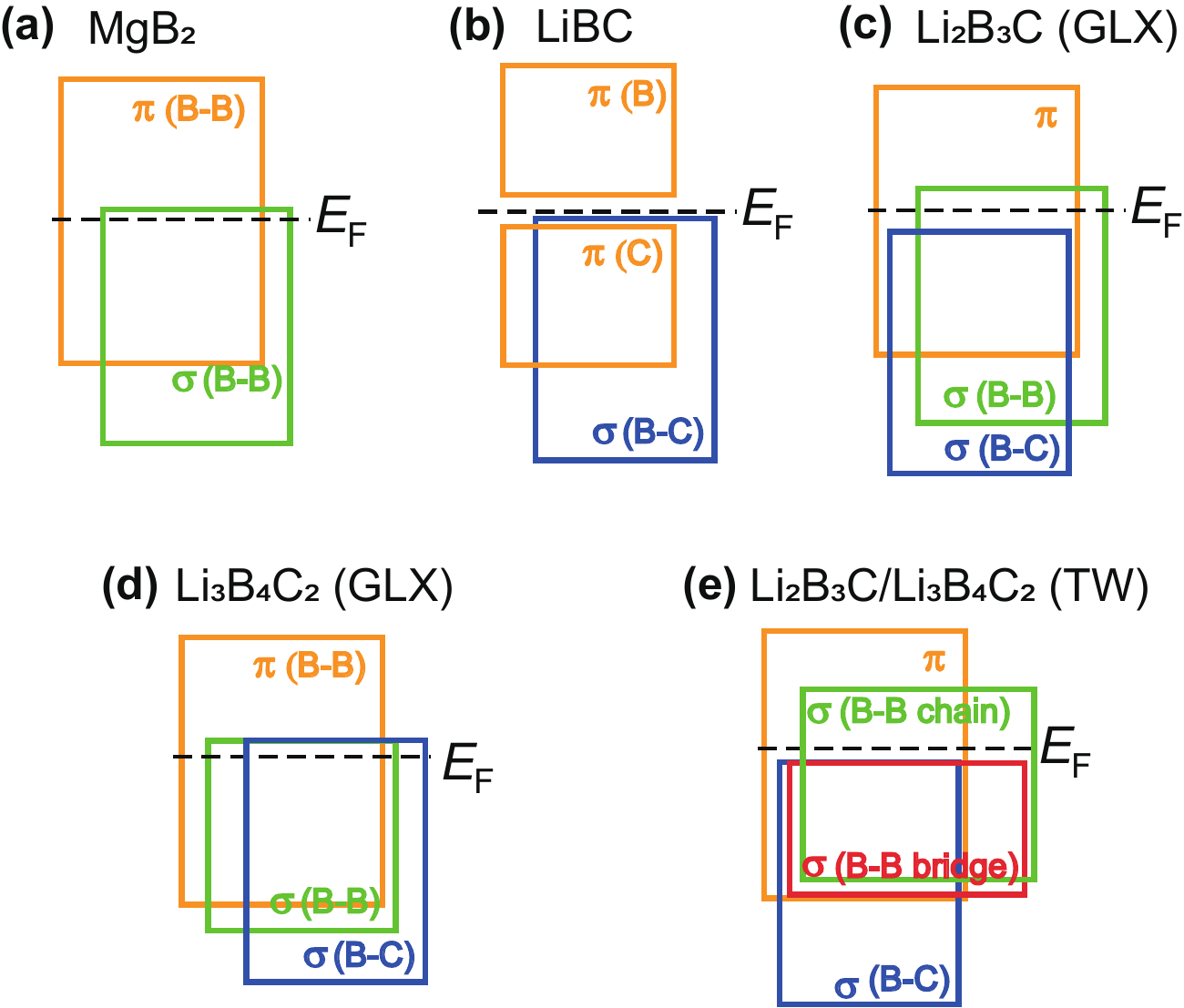}} \caption{Schematic band diagrams for (a) \ce{MgB2}, (b) \ce{LiBC}, (c) \ce{Li2B3C} (GLX), (d) \ce{Li3B4C2} (GLX), (e) \ce{Li2B3C}/\ce{Li3B4C2} (TW). 
		\label{MO} }
\end{figure}

Based on the analysis before, we can conclude simplified physical pictures for the electronic structures of these Li-B-C compounds and the prototype \ce{MgB2}, as diagrammatically sketched in Fig. \ref{MO}.

The crystal field effect of \ce{Mg^{2+}} ion gives rise to metallic $sp^2$ $\sigma$-bonding electrons in \ce{MgB2}. Holes from its $\pi$ bands are doped into its $\sigma$-bonding bands, resulting in the Fermi level being near the top of the $\sigma$-bonding bands, as shown in Fig. \ref{MO}(a). \ce{LiBC}, similar to BN, functions as an insulator, as evidenced in Fig. \ref{MO}(b). 

\ce{Li2B3C} (GLX) can be conceptualized as an amalgamation of LiBC and \ce{MgB2}. The electronic behavior near the Fermi level in this compound closely resembles that in \ce{MgB2} with more hole doping, as shown in Fig. \ref{MO}(c). In \ce{Li3B4C2} (GLX), the B-B and B-C $\sigma$-bonding bands are robustly hybridized and consequently indiscernible. Both sets of $\sigma$-bonding bands intersect the Fermi level, a phenomenon attributed to the crystal field effect and boron substitution.

In the case of \ce{Li2B3C} (TW) or \ce{Li3B4C2} (TW), only the $\sigma$-bonding bands from B-B zigzag chains and the $\pi$ bands intersect the Fermi level. Both compounds exhibit similar doping levels, with approximately 1/2 hole per $\sigma$-bonding WF from B-B zigzag chains. As discussed in our main text, we postulate that this principle is likely to be universally applicable across other \ce{Li_nB_{n+1}C_{n-1}} compounds.

\section{Wannierization of L\lowercase{i}B$_{n+1}$C$_{n-1}$ and M\lowercase{g}B$_2$}
\begin{figure}[htb]
	\centerline{\includegraphics[width=0.5\textwidth]{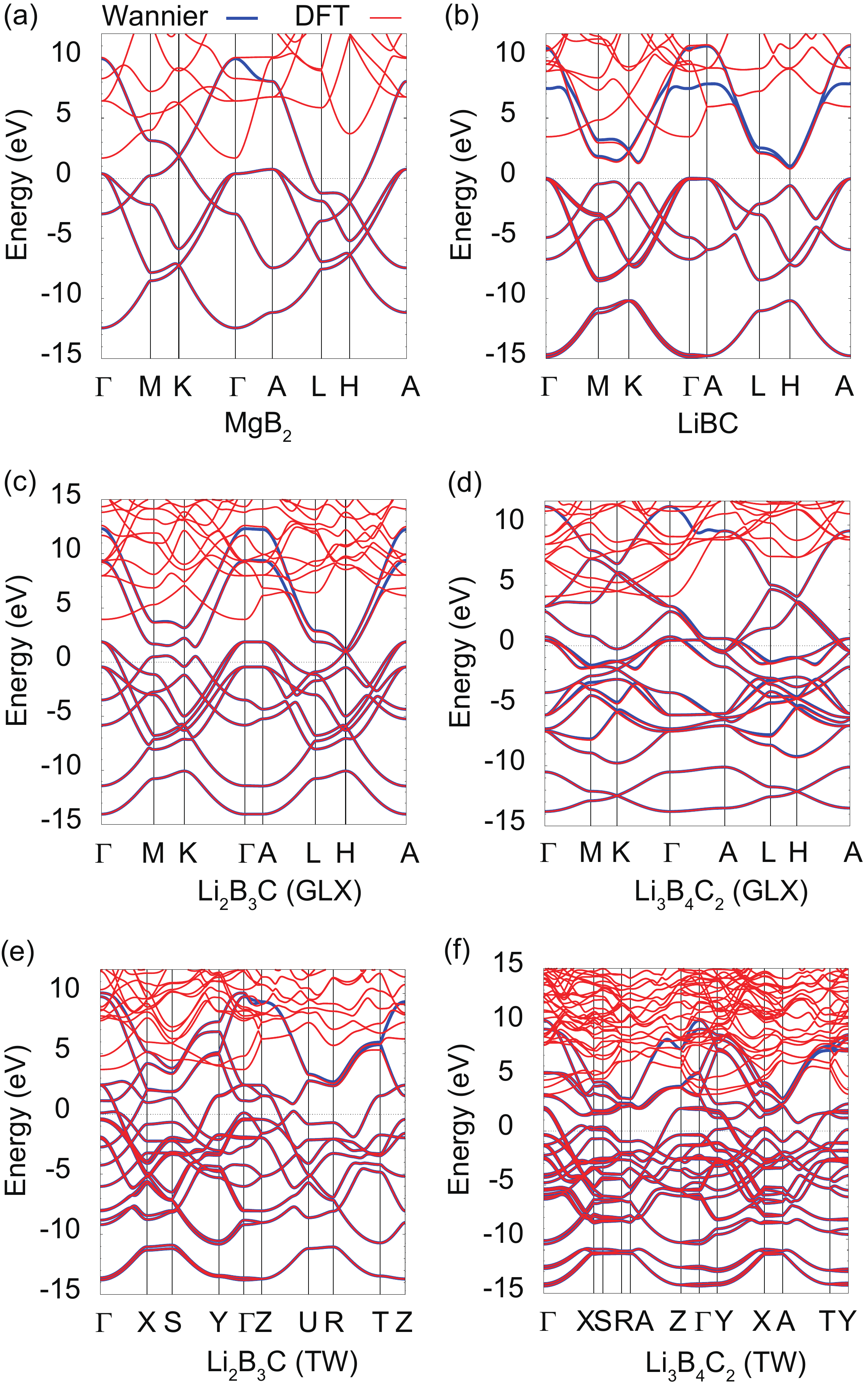}}
 \caption{Band structures of (a) \ce{MgB2}, (b) \ce{LiBC}, (c) \ce{Li2B3C} (GLX), (d) \ce{Li3B4C2} (GLX), (e) \ce{Li2B3C} (TW), (f) \ce{Li3B4C2} (TW). The red/blue lines represent DFT/Wannierization result, respectively.
		\label{wan-dft} }
\end{figure}

Our Wannierization results can perfectly reproduce the band structures of DFT calculations, as shown in Fig. \ref{wan-dft}. The computational details of our Wannierization are listed in TABLE. \ref{energy window}.
\begin{table}
\caption{\label{energy window}%
The energy window and frozen energy window for the disentanglement procedure in Wannierization. The Fermi level is set as 0. }
\begin{ruledtabular}
\begin{tabular}{ccc}
 Compound           & Energy window (eV)  & Frozen energy window (ev) \\
 \colrule
 \ce{MgB2}          & -21.72 $\sim$ 10.28 & -21.72 $\sim$ 0.00   \\
 \ce{LiBC}          & -19.66 $\sim$ 12.49 & -19.66 $\sim$ 0.00   \\
 \ce{Li2B3C} (TW)   & -16.04 $\sim$ 11.08 & -16.04 $\sim$ 0.00  \\
 \ce{Li2B3C} (GLX)  & -16.24 $\sim$ 13.06 & -16.24 $\sim$ 2.00  \\
 \ce{Li3B4C2} (TW)  & -16.70 $\sim$ 10.50 & -16.70 $\sim$ 1.30 \\
 \ce{Li3B4C2} (GLX) & -16.51 $\sim$ 11.59 & -16.51 $\sim$ 0.99 \\
\end{tabular}
\end{ruledtabular}
\end{table}

\clearpage

\section{details of the electronic structures of \ce{Li2B3C} (TW) under compressive strains}

\begin{figure}[htb]
 \centerline{\includegraphics[width=0.5\textwidth]{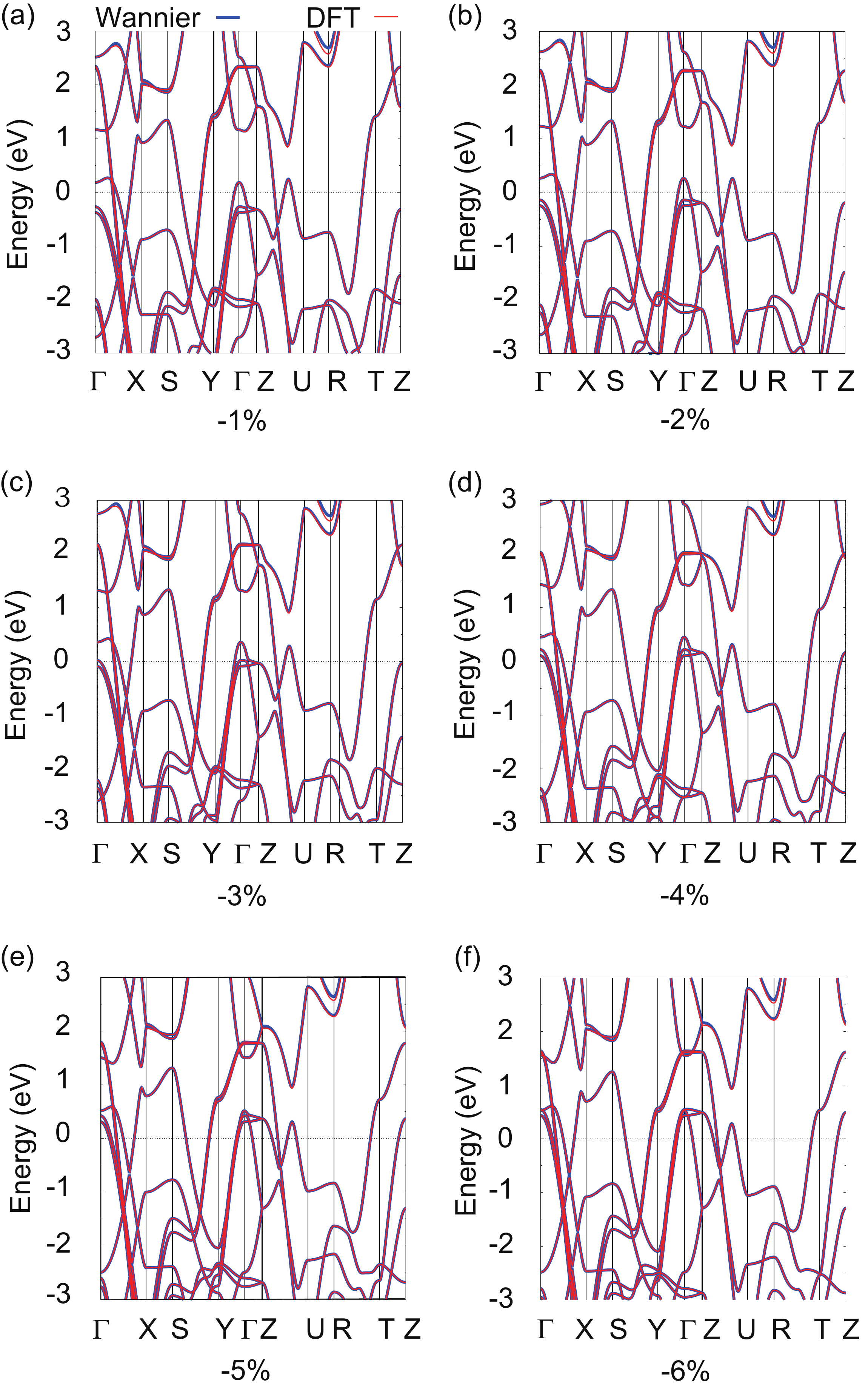}}
 \caption{Band structures \ce{Li2B3C} (TW) with different compressive $b$-axis strains. The red/blue lines represent DFT/Wannierization result, respectively.
		\label{strain-wan} }
\end{figure}

Our Wannierization results can perfectly reproduce the band structures of DFT calculations under compressive $b$-axis strains, as shown in Fig. \ref{strain-wan}. Moreover, we can see how bands evolve under greater compressive $b$-axis strain: the hole doping level of the bridge B-B $\sigma$-bonding bands increases, resulting in larger Fermi surfaces and a higher DOS of the bridge B-B $\sigma$-bonding WFs. $-4\%$ or more compressive $b$-axis strain gives rise to a sizable Fermi level composed by bridge B-B $\sigma$-bonding WFs with large deformation potentials, assuring the boost of $T_c$.

\clearpage

\end{document}